\documentclass[aps,twocolumn,english,superscriptaddress,citeautoscript,showkeys,preprintnumbers,amsmath,amssymb,floatfix,footinbib]{revtex4-2}
\usepackage{hyperref}
\hypersetup{colorlinks=true,linkcolor=red,citecolor=blue}
\usepackage{amsmath}
\usepackage{graphicx}
\usepackage{dcolumn}
\usepackage{xcolor}
\usepackage{float}
\usepackage{soul}

\mathchardef\mhyphen="2D

\begin{document}

	\title{A crossover from Kondo semiconductor to metallic antiferromagnet with $5d$-electron doping in  CeFe$_2$Al$_{10}$}
	\author{Rajesh Tripathi}
	\email{rajeshtripathi@jncasr.ac.in}
	\affiliation{ISIS Facility, STFC, Rutherford Appleton Laboratory, Chilton, Oxon OX11 0QX, United Kingdom}
	\affiliation{Jawaharlal Nehru Centre for Advanced Scientific Research, Jakkur, Bangalore 560064, India}
	
	\author{D. T. Adroja}
	\email{devashibhai.adroja@stfc.ac.uk}
	\affiliation{ISIS Facility, STFC, Rutherford Appleton Laboratory, Chilton, Oxon OX11 0QX, United Kingdom}
	\affiliation{Highly Correlated Matter Research Group, Physics Department, University of Johannesburg, Auckland Park 2006, South Africa}
	
	\author{M. R. Lees}
	\affiliation{Department of Physics, University of Warwick, Coventry CV4 7AL, United Kingdom}
	
	\author{ A. Sundaresan}
	\affiliation{Jawaharlal Nehru Centre for Advanced Scientific Research, Jakkur, Bangalore 560064, India}
	
	\author{S. Langridge}
	\affiliation{ISIS Facility, STFC, Rutherford Appleton Laboratory, Chilton, Oxon OX11 0QX, United Kingdom}

	\author{A. Bhattacharyya}
	\affiliation{Department of Physics, Ramakrishna Mission Vivekananda Educational and Research Institute, Belur Math, Howrah 711202, West Bengal, India}

	\author{V. K. Anand}
	\affiliation{\mbox{Helmholtz-Zentrum Berlin f\"{u}r Materialien und Energie GmbH, Hahn-Meitner Platz 1, D-14109 Berlin, Germany}}
	
	\affiliation{Department of Physics, University of Petroleum and Energy Studies, Dehradun, Uttarakhand, 248007, India}

	\author{D. Khalyavin}
	\affiliation{ISIS Facility, STFC, Rutherford Appleton Laboratory, Chilton, Oxon OX11 0QX, United Kingdom}

	\author{ J. Sannigrahi}
	\affiliation{ISIS Facility, STFC, Rutherford Appleton Laboratory, Chilton, Oxon OX11 0QX, United Kingdom}

	\author{G. Cibin}
	\affiliation{Diamond Light Source, Harwell Science and Innovation Campus, Didcot, Oxfordshire OX11 0DE, United Kingdom}
	\author{A. D. Hillier}
	\affiliation{ISIS Facility, STFC, Rutherford Appleton Laboratory, Chilton, Oxon OX11 0QX, United Kingdom}
	\author{R. I. Smith}
	\affiliation{ISIS Facility, STFC, Rutherford Appleton Laboratory, Chilton, Oxon OX11 0QX, United Kingdom}
	\author{H. C. Walker}
	\affiliation{ISIS Facility, STFC, Rutherford Appleton Laboratory, Chilton, Oxon OX11 0QX, United Kingdom}
	\author{Y. Muro}
	\affiliation{Liberal Arts and Sciences, Faculty of Engineering, Toyama Prefectural University, Imizu 939-0398, Japan}
	
	\author{T. Takabatake}
	\affiliation{Quantum Matter Program, Graduate School of Advanced Science and Engineering, 
		Hiroshima University, Higashi-Hiroshima 739-8530, Japan}

	\date{\today}

	\begin{abstract}
		
		We report a systematic study of the $5d$-electron-doped system Ce(Fe$_{1-x}$Ir$_x$)$_2$Al$_{10}$ ($0 \leq x \leq 0.15$). With increasing $x$, the orthorhombic $b$~axis decreases slightly while accompanying changes in $a$ and $c$ leave the unit cell volume almost unchanged. Inelastic neutron scattering, along with thermal and transport measurements, reveal that for the Kondo semiconductor CeFe$_2$Al$_{10}$, the low-temperature energy gap which is proposed to be a consequence of strong $c \mhyphen f$ hybridization, is suppressed by a small amount of Ir substitution for Fe, and that the system adopts a metallic ground state with an increase in the density of states at the Fermi level. The charge or transport gap collapses (at $x=$~0.04) faster than the spin gap with Ir substitution. Magnetic susceptibility, heat capacity, and muon spin relaxation measurements demonstrate that the system undergoes long-range antiferromagnetic order below a N\'eel temperature, $T_{\mathrm{N}}$, of 3.1(2)~K for $x = 0.15$. The ordered moment is estimated to be smaller than 0.07(1)~$\mu_\mathrm{B}$/Ce although the trivalent state of Ce is confirmed by Ce L$_3$-edge x-ray absorption near edge spectroscopy. It is suggested that the $c \mhyphen f$ hybridization gap, which plays an important role in the unusually high ordering temperatures observed in Ce$T_2$Al$_{10}$ ($T$ = Ru and Os), may not be necessary for the onset of magnetic order with a low $T_{\mathrm{N}}$ seen here in Ce(Fe$_{1-x}$Ir$_x$)$_2$Al$_{10}$.

		
	\end{abstract}
	
	\maketitle
	
	\section{INTRODUCTION}
	\label{Intro}
	
	Cerium-based intermetallic systems Ce$T_2$Al$_{10}$ ($T = $~Os, Ru, or Fe) with an orthorhombic YbFe$_2$Al$_{10}$-type crystal structure (space group $Cmcm$) have attracted considerable interest in recent years~\cite{A705854C,200916000,STRYDOM20092981,083707,PhysRevB.82.100404,PhysRevB.82.100405,PhysRevB.84.165125,PhysRevLett.109.267208,JOUR,doi:10.7566/JPSJ.83.094717,PhysRevB.92.201113,PhysRevB.94.165137,PhysRevB.95.035144,PhysRevB.97.155106,PhysRevB.103.045101,PhysRevLett.106.056404}. In these materials, a hybridization gap opens in the vicinity of the Fermi level owing to the strong coupling between the localized $4f$ electrons and itinerant conduction electrons ($c \mhyphen f$ hybridization) at low temperatures~\cite{PhysRevB.84.165125,doi:10.7566/JPSJ.83.094717,PhysRevB.92.201113}. Because of the hybridization gap, these materials display semiconducting or semi-metallic behavior in the electrical resistivity, and are known as ``Kondo semiconductors''. In general, it is believed that Kondo semiconductors have non-magnetic ground states e.g., SmB$_6$~\cite{MOSHCHALKOV1985289}, YbB$_{12}$~\cite{KASAYA1983437}, Ce$_3$Bi$_4$Pt$_3$~\cite{EKINO1998379}, CeRhSb \cite{PhysRevB.43.6277,PhysRevLett.75.4262}. The Kondo semiconductors CeRu$_2$Al$_{10}$ and CeOs$_2$Al$_{10}$ on the other hand, were reported to harbor an antiferromagnetic (AFM) ground state with $T_{\mathrm{N}} = 27.3$ and 28.7~K, respectively, with a pseudo-gap at a temperature slightly higher than $T_{\mathrm{N}}$~\cite{STRYDOM20092981,200916000,doi:10.1143/JPSJ.78.123713}. The magnetic ordering temperature was unexpectedly high, despite the large separation ($\sim5.2$~\AA) between the Ce ions. Their Gd counterparts have relatively lower ordering temperatures $T_{\mathrm{N}} = 17.5$~K~\cite{doi:10.7566/JPSJ.86.094709} and 18~K~\cite{doi:10.1143/JPSJS.80SA.SA021}, respectively. Moreover, the small Ce ordered moments of $\sim$ 0.3–0.4 $\mu_{\mathrm{B}}$, are aligned along the $c$ axis despite the large uniaxial anisotropy in the magnetic susceptibility $\left(\chi_a \gg \chi_c > \chi_b\right)$ in the paramagnetic state~\cite{PhysRevB.82.100405,PhysRevB.84.233202,doi:10.1143/JPSJ.79.083701}. These observations challenge the conventional indirect Ruderman–Kittel–Kasuya–Yosida (RKKY) exchange interaction and crystalline electric field (CEF) models and the ordering mechanism is still under discussion. An optical conductivity measurement on CeOs$_2$Al$_{10}$ has revealed that a charge-density-wave like instability which develops along the $b$ axis at temperatures slightly higher than $T_{\mathrm{N}}$ induces unconventional AFM ordering~\cite{PhysRevLett.106.056404}. This mechanism contrasts with conventional RKKY dominated ordering as in the case of Gd$T_2$Al$_{10}$~\cite{PhysRevB.91.241120,PhysRevB.84.165125} and NdOs$_2$Al$_{10}$~\cite{10.1088/1361-648X}. The orthorhombic crystalline-electric field experienced by the $f$-electrons was also proposed to be an important driver for the strong magnetic anisotropy and the small ordered magnetic moments in Ce$T_2$Al$_{10}$~\cite{200916000,PhysRevB.81.214401,JOUR,PhysRevB.87.125119,PhysRevB.86.081105}. In the case of CeFe$_2$Al$_{10}$, the presence of strong $c\mhyphen f$ hybridization leads to a screening of the localized moments by the conduction electron spins, and hence the compound is categorized as a valence fluctuating material~\cite{JOUR,PhysRevB.87.125119,doi:10.7566/JPSJ.83.084708}. Due to the strong hybridization, CeFe$_2$Al$_{10}$ is expected to have the largest real or pseudo-gap feature among the Ce$T_2$Al$_{10}$ materials~\cite{PhysRevB.92.201113,TransTech}. A pseudo-gap of 12~meV near the Fermi level ($E_{\mathrm{F}}$) at 10~K was identified by high-resolution photoemission spectroscopy~\cite{doi:10.7566/JPSJ.83.094717}. Inelastic neutron scattering (INS) measurements also confirmed the spin gap or ``Kondo gap'' of 12.5~meV~\cite{PhysRevB.87.224415} and further confirmation of the $q$-dependence of the gap was reported through INS studies using single-crystals~\cite{Mignot2014, TransTech}. These results suggested a possible Kondo semiconducting nature for CeFe$_2$Al$_{10}$ similar to Ce$_3$Bi$_4$Pt$_3$~\cite {PhysRevB.44.6832} and SmB$_6$~\cite {PhysRevLett.114.036401}.
	
	It can be inferred from these findings that the strength of the $c \mhyphen f$ coupling plays an important role in the unusual magnetic ordering in Ce$T_2$Al$_{10}$. The $c \mhyphen f$ coupling can be controlled by the application of a magnetic field or pressure, as well as by atomic substitution. For example, previous experimental studies have shown that the magnetic properties are profoundly affected by $T$-site substitution. The substitution of Ir (electrons) or Re (holes) on the $T$-site in CeOs$_2$Al$_{10}$ has a significant effect on the magnetic and other physical properties. In Ce(Os$_{1-x}$Ir$_x$)$_2$Al$_{10}$, the Ir substitution drives the system into a more localized (Ce$^{3+}$) regime with an ordered moment that increases to 0.92 $\mu_{\mathrm{B}}$ for $x = 0.08$, along with a spin reorientation from the $c$ to the easy $a$ axis~\cite {PhysRevB.88.060403}. The $T_{\mathrm{N}}$, however, decreases from 28.5~K for $x = 0$ to 21~K for $x = 0.08$ and the spin-gap excitations near 11~meV are considerably suppressed~\cite{PhysRevB.90.174422}. Re substitution, on the other hand, leads to an enhancement of the $c\mhyphen f$ hybridization, which causes a reduction of $T_{\mathrm{N}} = 21$~K and a reduced ordered moment of 0.18(1)$\mu_B$ for $x = 0.03$ \cite{PhysRevB.89.064422}. However, the direction of the ordered moment remains unchanged and magnetic order is suppressed for more than 5\% Re-doping~\cite{PhysRevB.89.094404,PhysRevB.90.174422}. A similar chemical doping effect has also been observed in the Rh/Fe-substituted CeRu$_2$Al$_{10}$~\cite {doi:10.7566/JPSJ.82.093702,doi:10.7566/JPSJ.83.104707,JPCMCS.273.012046}. It was seen that the hybridization gap is strongly suppressed by Rh doping despite the more localized nature of the $4f$ state, while Fe-doping strengthens the $c \mhyphen f$ hybridization, leading to a delocalization of the $4f$ state~\cite{PhysRevB.91.241120,PhysRevB.94.165137,PhysRevB.87.224415,PhysRevB.89.125108}. Since the suppression of $T_{\mathrm{N}}$ in these substituted systems seems to be correlated with the suppression of the hybridization gap, it was proposed that the hybridization gap is necessary for the unusual AFM order with a high $T_{\mathrm{N}}$~\cite{PhysRevB.89.094404}.
	
	The strength of $c \mhyphen f$ hybridization in most of the Ce-based compounds is increased by hydrostatic pressure, $P$, leading to a suppression of the AFM ordered state. While, in Ce$T_2$Al$_{10}$ for $T=$~Ru and Os, upon increasing $P$, $T_{\mathrm{N}}$ first increases to a maximum of 32 and 39~K and then rapidly falls to zero at 4.8 and 2.5~GPa, respectively~\cite{200916000,doi:10.1143/JPSJ.80.064709}. Since the semiconducting behavior of resistivity remains in the AFM ordered state and then disappears along with $T_{\mathrm{N}}$, the hybridization gap was thought to be essential for the unusual AFM order. Another puzzling feature is that application of $P \parallel b$ strongly increases $T_{\mathrm{N}}$ in Ce$T_2$Al$_{10}$ ($T = $~Ru, Os)~\cite{PhysRevB.96.245130}.

	Rh substitution in CeFe$_2$Al$_{10}$ lead to a similar localization of the $4f$ state, but does not induce magnetic ordering for up to 20\% Rh substitution down to 2~K. However, a closing of the Kondo-gap and the onset of a metallic ground state are observed~\cite{PhysRevB.92.235154,PhysRevB.102.024438}. 
	
	Given that the Ce$T_2$Al$_{10}$ family is highly sensitive to both pressure and the nature of the $d$-electrons [$3d$ (Fe), and $4d$(Ru)], it is interesting to investigate the effects of adding $5d$ electrons in CeFe$_2$Al$_{10}$, e.g. Ir substitution, which has stronger spin-orbit coupling and a larger ionic radius than Fe and Ru~\cite{Shannon:a12967}. Our motivation for investigating the Ce(Fe$_{1-x}$Ir$_x$)$_2$Al$_{10}$ system is therefore to study the relationship between the anisotropic $c \mhyphen f$ hybridization, the spin gap, the Kondo semiconducting behavior, and the onset of magnetic order.

	Here we report the results of the macroscopic (thermal, transport, and magnetic) and microscopic (muon spin relaxation, elastic and inelastic neutron scattering) measurements on both polycrystalline ($x = 0$, 0.04, 0.08, 0.15) and single crystal ($x = 0.15$) samples of Ce(Fe$_{1-x}$Ir$_x$)$_2$Al$_{10}$. These results reveal a long-range ordered AFM state at around 3.1(2)~K for $x = 0.15$. The substitution of Ir leads to a decrease in the $b$ axis lattice parameter and increases in both the $a$ and $c$ parameters. We therefore propose that the reduced lattice parameter $b$, together with an increase in the number of $5d$ electrons, weakens the hybridization between the $4f$ and the conduction electrons which results in a rapid collapse of the spin/charge gap and an increase of the density of states (DOS) at the Fermi level. As a result, an AFM metallic ground state is realized in this system.

	\section{EXPERIMENTAL METHODS}
	\label{expt}
	
	Polycrystalline samples of Ce(Fe$_{1-x}$Ir$_x$)$_2$Al$_{10}$ for $0 \leq x \leq 0.15$ were prepared by arc melting stoichiometric quantities of the constituent elements under a high-purity Ar atmosphere. The samples were heat treated in evacuated quartz tubes at 1000~$^\circ$C for one week. Single crystals for $x = 0.15$ were prepared by an Al self-flux method using the same procedure as given in Ref.~\onlinecite{083707}. Powder neutron diffraction (PND) measurements at room temperature on the GEM diffractometer at the ISIS Facility, Rutherford Appleton Laboratory, UK were used to determine the phase purity and crystal structure of the polycrystalline samples, and Laue x-ray diffraction was used to check the quality of the single crystal. A scanning electron microscope equipped with energy dispersive x-ray analysis was used to check the homogeneity and composition of the single crystal and the results are presented in the Supplemental Material~\cite{SM}. 
	
	DC electrical resistivity, $\rho$, measurements were performed as a function of temperature down to 2~K in a Quantum Design (QD) Physical Property Measurement System (PPMS) using a standard four-probe technique with a measuring current of 10~mA. Thermoelectric power, $S$, was measured with the thermal transport option of the PPMS. Heat capacity was measured dow to 0.4~K using a helium-3 option by the relaxation method in a PPMS. Magnetization, $M$, measurements were performed in the temperature range 2 - 300~K using a vibrating sample magnetometer option of a PPMS.

	Muon spin relaxation ($\mu$SR) measurements were performed in zero-field (ZF) using the MuSR spectrometer at the ISIS Facility. The powdered sample with $x =$~0.15 (approximately 4~grams) was mounted on a 99.999\% pure silver plate using diluted General Electric varnish and then covered with a thin silver foil. The ZF-$\mu$SR spectra were recorded at several temperatures between 1.2 and 5~K. See Refs.~\onlinecite{PhysRevB.87.224415,PhysRevB.99.224424} for a detailed description of the experimental technique. Powder neutron diffraction measurements at 1.5 and 5~K on a sample with $x =$~0.15 were performed using the WISH diffractometer at the ISIS Facility. For both the GEM and WISH data, the nuclear and magnetic intensities were refined by the Rietveld technique using the FullProf software suite~\cite {https://www.ill.eu/sites/fullprof/}. Inelastic neutron scattering measurements at 5~K for $x = 0.04$ and 0.08 were performed using the MERLIN time-of-flight chopper spectrometer at the ISIS Facility. Ce L$_3$-edge x-ray absorption near-edge structure (XANES) spectra for $x = 0.04$ and 0.15 were measured in transmission mode (at 7~K) using the general purpose x-ray absorption spectroscopy beamline B18 at the Diamond Light Source, UK. The samples were prepared by grinding the polycrystalline material into fine powder, mixing it with cellulose, and pressing the mixture into pellets.
	
	\begin{figure}
		\includegraphics[width=7.5cm, keepaspectratio]{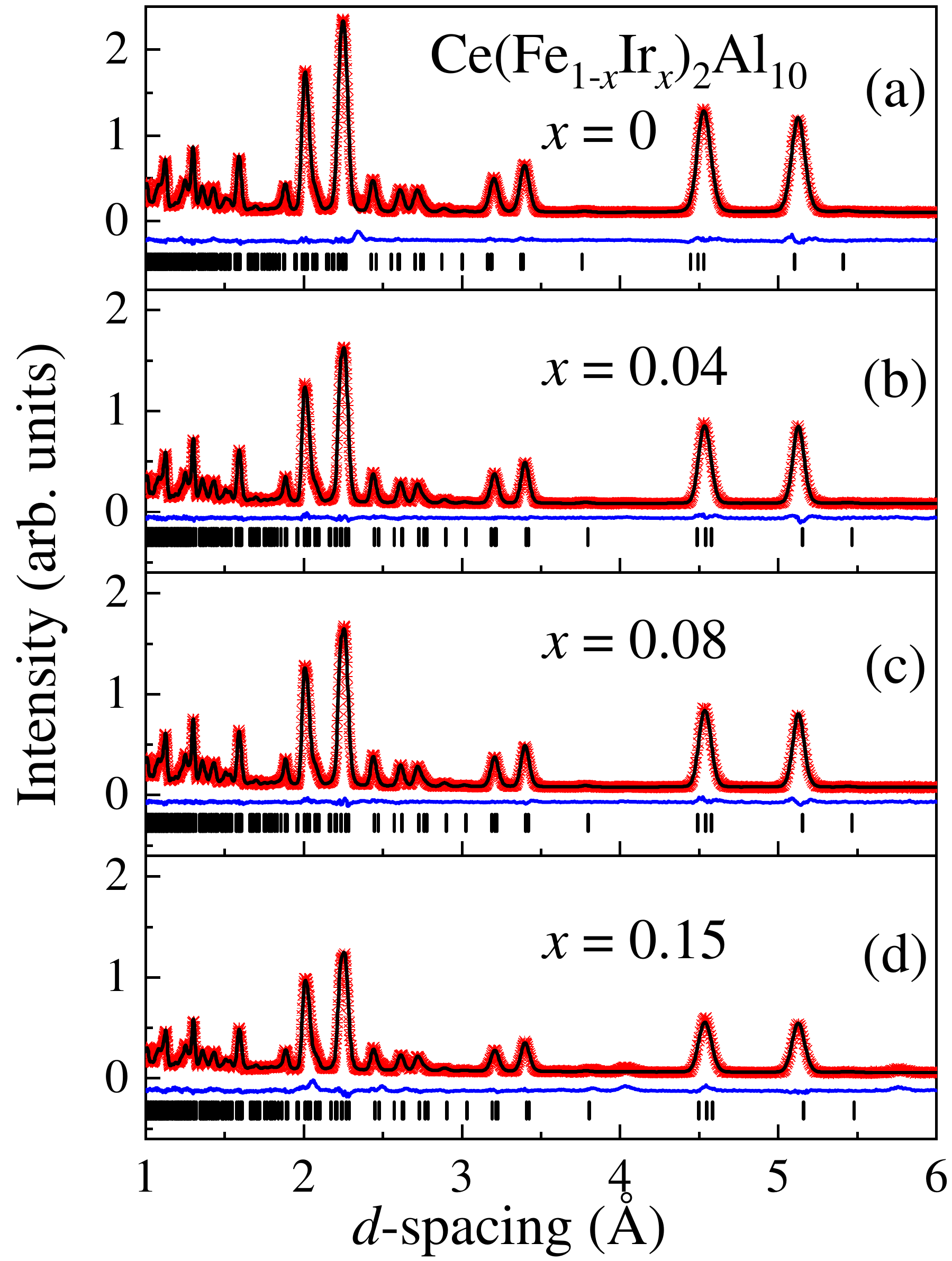}
		\caption{Powder neutron diffraction patterns of Ce(Fe$_{1-x}$Ir$_x$)$_2$Al$_{10}$ $\left(0 \leq x \leq 0.15\right)$ obtained at room temperature using the GEM diffractometer. The observed and calculated intensities, and difference are plotted as red symbols, a solid black line, and a solid blue line, respectively. The vertical tick marks show the positions of the Bragg reflections}
		\label{PNDdata}
	\end{figure}

	\begin{figure}
		\includegraphics[width=8.5cm, keepaspectratio]{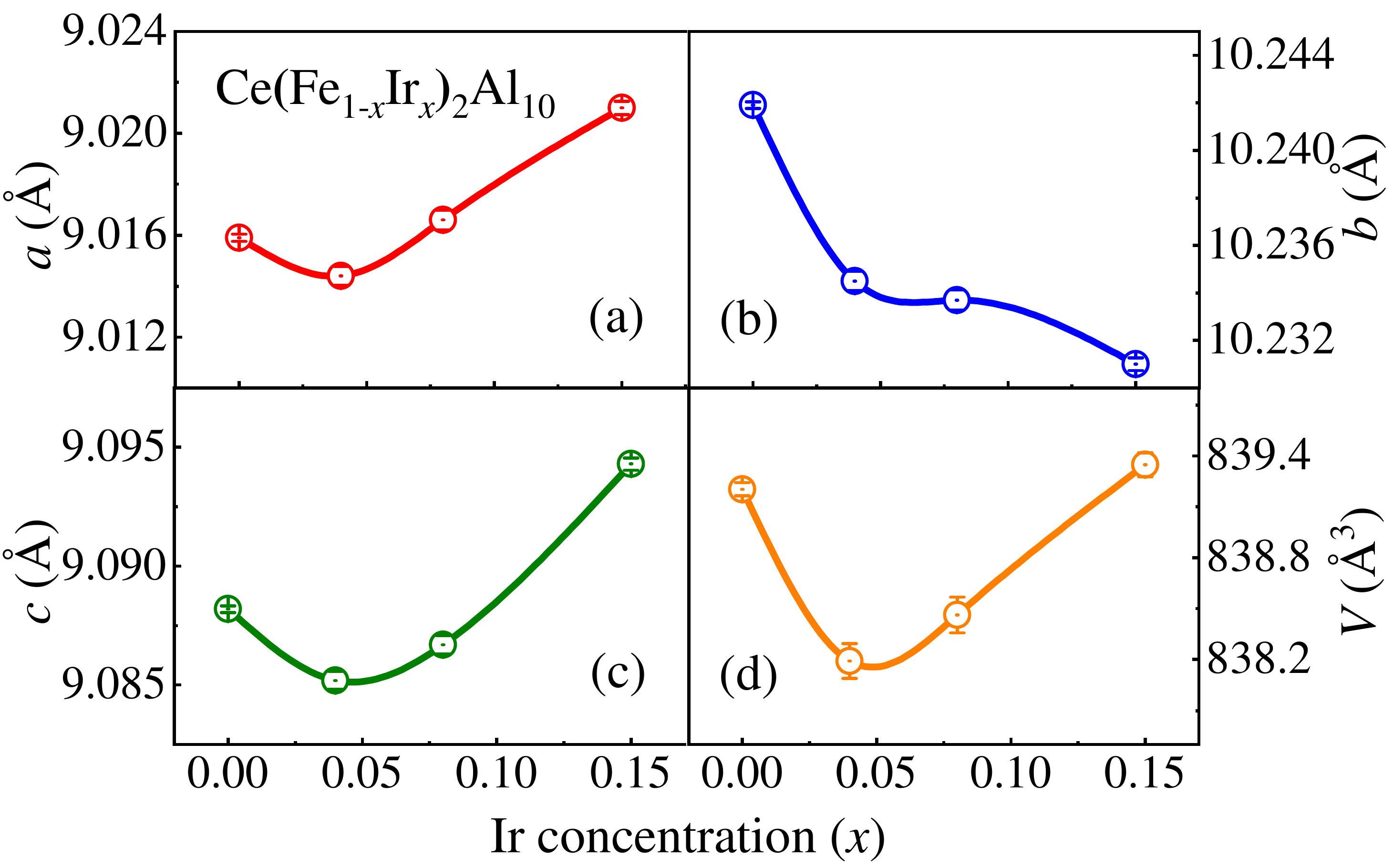}
		\caption{Variation of the orthorhombic lattice parameters $a$, $b$, and $c$, and unit-cell volume, $V$ as a function of Ir concentration $x$ for Ce(Fe$_{1-x}$Ir$_x$)$_2$Al$_{10}$.}
		\label{LatticeParameters}
	\end{figure}

	\section{RESULTS AND DISCUSSION}
	
	\subsection{Crystal Structure}
	\label{structure}
	Figure~\ref{PNDdata} shows the powder neutron diffraction patterns of Ce(Fe$_{1-x}$Ir$_x$)$_2$Al$_{10}$ for $0 \leq x \leq 0.15$ at room temperature. The  data show that all of the Ir-doped materials adopt the orthorhombic YbFe$_2$Al$_{10}$-type structure with space group $Cmcm$. The crystal structures were refined by the Rietveld method~\cite {https://www.ill.eu/sites/fullprof/} and the lattice parameters and unit cell volume as a function of the Ir concentration $x$ shown in Fig.~\ref{LatticeParameters}. The lattice parameters exhibit anomalous changes. The $b$~axis decreases with increasing $x$ up to $x = 0.15$, while $a$ and $c$ have minimum for $x = 0.04$ and then increase with increasing $x$. As shown in Fig.~\ref{LatticeParameters}(d), the unit cell volume for $x=$~0.15 is almost identical to that for $x =$~0. This suggests that the overall chemical pressure effect will be negligible.
	
	It is important to note that the lattice parameters in Ce(Fe$_{1-x}$Rh$_x$)$_2$Al$_{10}$ isotropically increase by 0.3$\mhyphen$0.4\% as $x$ is increased to 0.2, leading to a negative pressure effect~\cite{PhysRevB.92.235154}. Upon applying pressure, the lattice parameters of Ce$T_2$Al$_{10}$ ($T = $~Fe, Ru, Os) all decrease monotonically but the response is anisotropic~\cite{doi:10.7566/JPSJ.85.044601}. A decrease in the $b$ axis lattice parameter under pressure leads to a concomitant increase in $T_{\mathrm{N}}$~\cite{PhysRevB.96.245130, doi:10.7566/JPSJ.85.044601}. We, therefore, anticipate a difference in the properties of Ir and Rh-doped CeFe$_2$Al$_{10}$, due to the different behavior of the $b$~axis lattice parameter with chemical substitution.

	\subsection{Electrical resistivity and thermoelectric power}
	\label{transport}
	
	\begin{figure}
		\includegraphics[width=7.5cm, keepaspectratio]{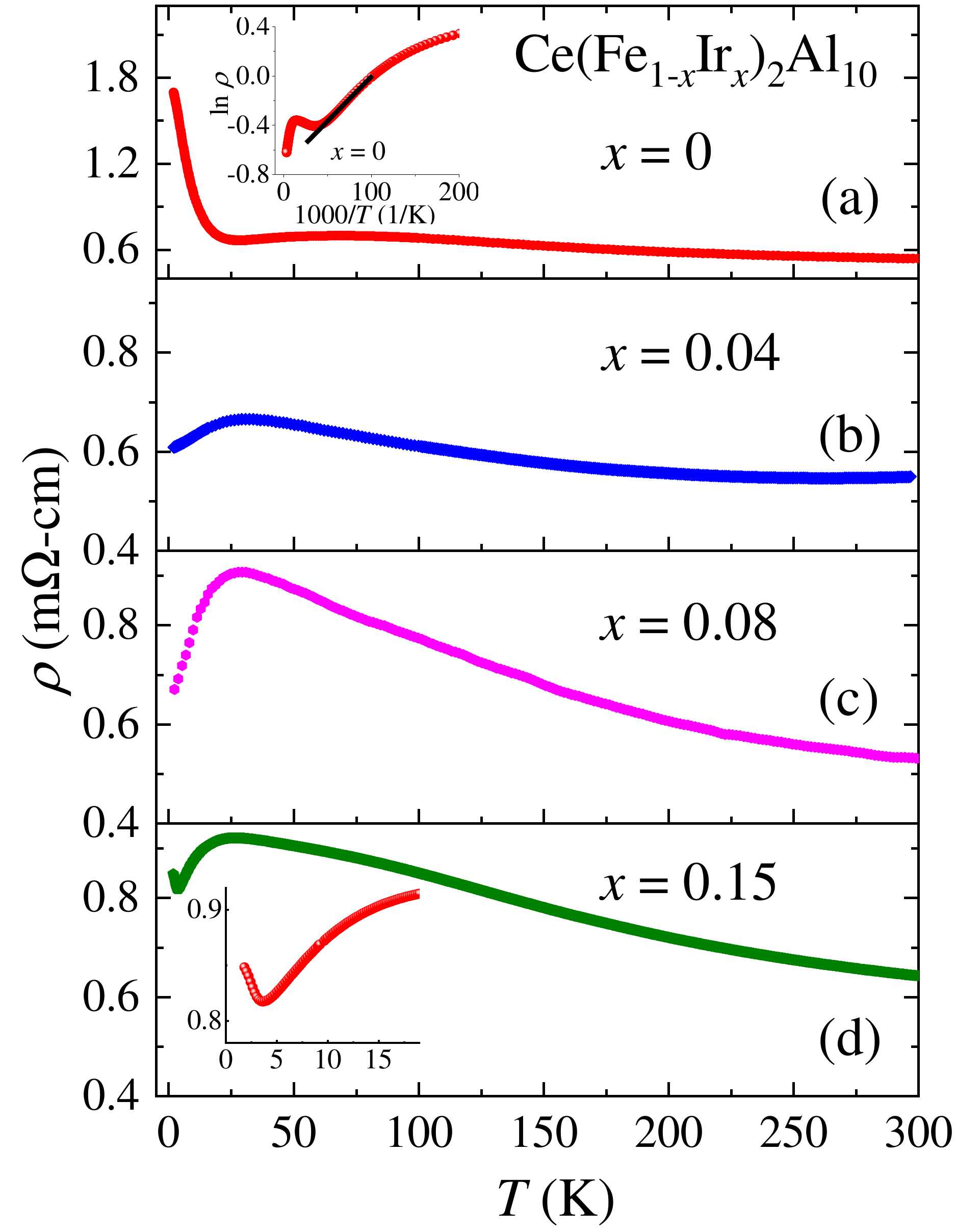}
		\caption{Electrical resistivity vs temperature for Ce(Fe$_{1-x}$Ir$_x$)$_2$Al$_{10}$ $\left(0 \leq x \leq 0.15\right)$. The inset in (a) shows the thermal activation behavior of the resistivity, while the inset in (d) shows an enlarged view of the low temperature upturn of $\rho(T)$ for $x = 0.15$.}
		\label{resistivity}
	\end{figure}
	
	Figures~\ref{resistivity}(a) -~\ref{resistivity}(d) show the temperature dependence of $\rho$ for Ce(Fe$_{1-x}$Ir$_x$)$_2$Al$_{10}$ ($x = 0$ 0.04, 0.08, and 0.15). The $\rho(T)$ curve for $x = 0$ gradually increases with decreasing temperature with a broad hump at $T_0 \sim 75$~K and a sharp upturn below $T = 20$~K, consistent with previous reports~\cite{083707}. The $-\log T$ behavior above $T_0$ can be ascribed to the Kondo scattering on the crystal field excited state. The Kondo semiconducting behavior at $T<20$~K results in a linear variation in $\ln \rho$ versus $1000/T$ as shown in the inset of Fig.~\ref{resistivity}(a). The low-temperature behavior is well described by the activation law $\rho(T)=\rho_{0} \exp \left(\Delta / 2 k_{\mathrm{B}} T\right)$. A least-squares fit to the $\rho\left(T\right)$ data between 10 and 20~K gives an energy gap $\Delta/k_{\mathrm{B}} = 15$~K, consistent with previous reports~\cite{PhysRevB.92.201113,083707}.

	\begin{figure}
		\includegraphics[width=8.5cm, keepaspectratio]{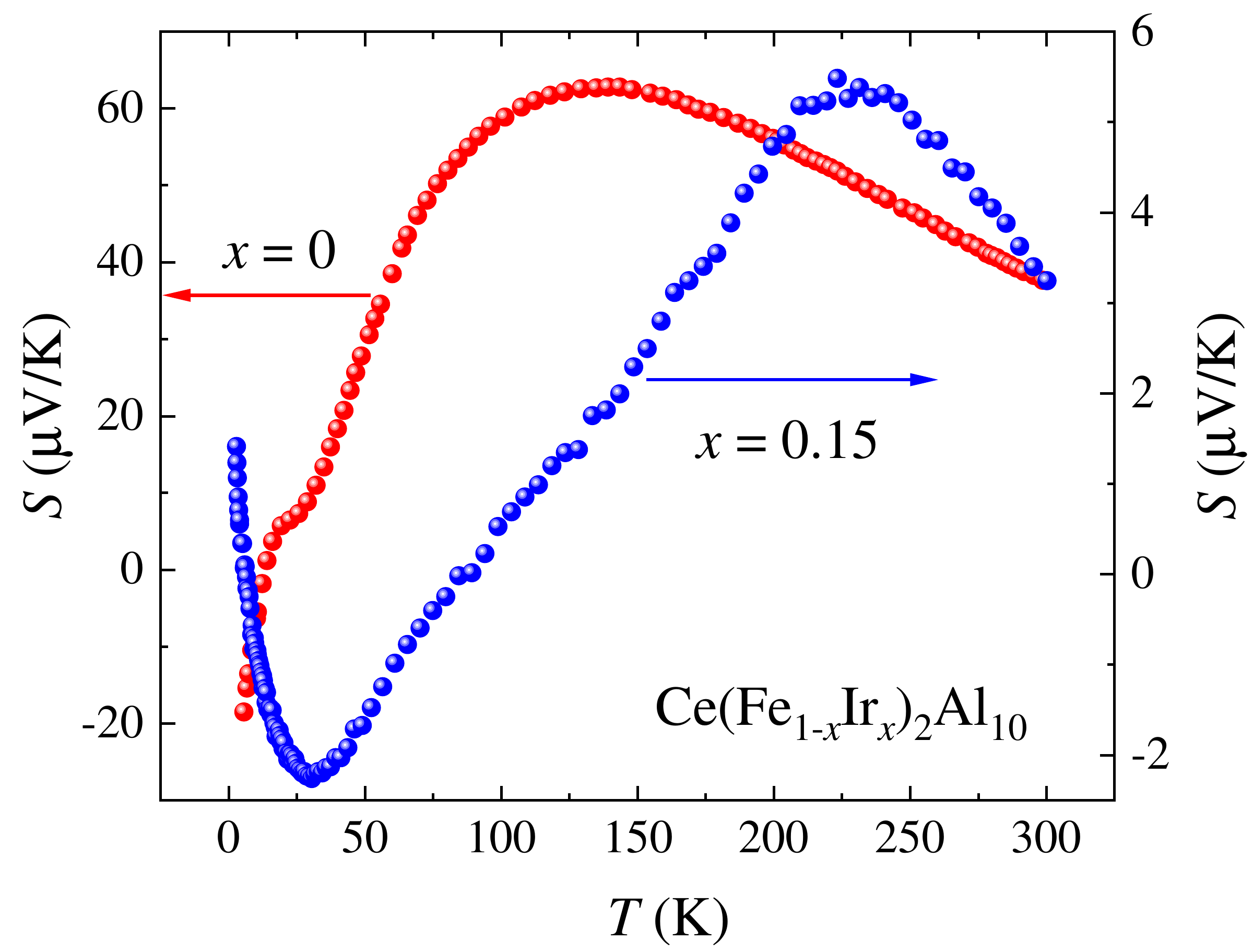}
		\caption{ Thermoelectric power data for Ce(Fe$_{1-x}$Ir$_{x}$)$_2$Al$_{10}$ with $x = 0$ (left axis) from Ref.~\cite{083707} and $x = 0.15$ (right axis).}
		\label{thermopower}
	\end{figure}
	
	A small amount of Ir substitution abruptly suppresses the Kondo semiconducting behavior below $T = 20$~K without any significant change in the room temperature resistivity. In fact, $\rho\left(T\right)$ of Ce(Fe$_{1-x}$Ir$_x$)$_2$Al$_{10}$ ($x = 0.04$, 0.08, 0.15) becomes metallic at low temperature. This strongly suggests that the Kondo semiconducting behavior in $\rho$(T) is suppressed by Ir (or electron) doping, and a metallic behavior is realized. The low-temperature maximum in $\rho$ at $T_0 \sim 30$~K for $x = 0.04$ is attributed to the onset of Kondo coherence. We note that this maximum shifts towards lower temperatures with increasing $x$. A similar doping effect on $\rho$(T) has been reported in CeOs$_2$Al$_{10}$ where the low-temperature upturn is strongly suppressed, and a metallic ground state has been realized for both electron (Ir/Rh) and hole (Re)-doping~\cite{PhysRevB.89.094404}. Therefore, the metallization in Fe/Os-based Ce$T_2$Al$_{10}$ materials seems to be favored by both electron and hole doping. A similar feature in $\rho\left(T\right)$ is also seen for Ce$T_2$Al$_{10}$ ($T = $~Ru and Os) under hydrostatic pressure~\cite{PhysRevB.96.245130}. The inset of Fig.~\ref{resistivity}(d) shows that for $x = 0.15$ a small upturn in $\rho\left(T\right)$ appears at 4~K, below the onset of the Kondo coherence (at 20~K). We attribute this behavior to the formation of a magnetic superzone gap associated with long-range magnetic ordering observed near 3.1(2)~K in the heat capacity and magnetic susceptibility discussed in the Sections~\ref{HeatCapacity} and ~\ref{MagSusc} below.
	
	The temperature dependence of the thermoelectric power, $S(T)$, for Ce(Fe$_{1-x}$Ir$_{x}$)$_2$Al$_{10}$ with $x = 0.15$ along with $S(T)$ for $x = 0$, obtained from Ref.~\onlinecite{083707} for comparison, are shown in Fig.~\ref{thermopower}. For CeFe$_2$Al$_{10}$, $S(T)$ is positive and significantly enhanced, which is typical of intermediate valence systems like CeNi$_2$Si$_2$~\cite{SAMPATHKUMARAN198971,KOTERLYN2005231} and CePd$_3$~\cite{JACCARD1987572}. In contrast, a very strong suppression of $S(T)$ is seen for $x = 0.15$. A broad maximum around 90~K is followed by a sign change at 34~K and a minimum of -2~$\mu$V/K. The overall $S(T)$ behavior is very similar to that observed for the AFM Kondo lattice compound CePdGa with $T_{\mathrm{N}} = 1.8$~K~\cite{ADROJA1994169, CePdGaTN}. The negative peak in $S(T)$ below 30~K is attributed to Kondo coherence~\cite{Fischer}.
	
	While $\rho(T)$ of Ce(Fe$_{0.85}$Ir$_{0.15}$)$_2$Al$_{10}$ exhibits a clear anomaly associated with the superzone gap formation indicating an incommensurate AFM ordering, there is no clear indication of magnetic ordering in $S(T)$ except a crossover from negative to positive values of $S$ that occurs near 7~K, at a temperature somewhat higher than the long-range AFM ordering. However, at approximately 3~K, both the temperature dependent $\chi(T)$ and $C(T)$ show a clear downward kink as expected for AFM ordering, as discussed in the following sections.

	\subsection{Heat Capacity}
	\label{HeatCapacity}
	
	\begin{figure}
		\includegraphics[width=8.5cm, keepaspectratio]{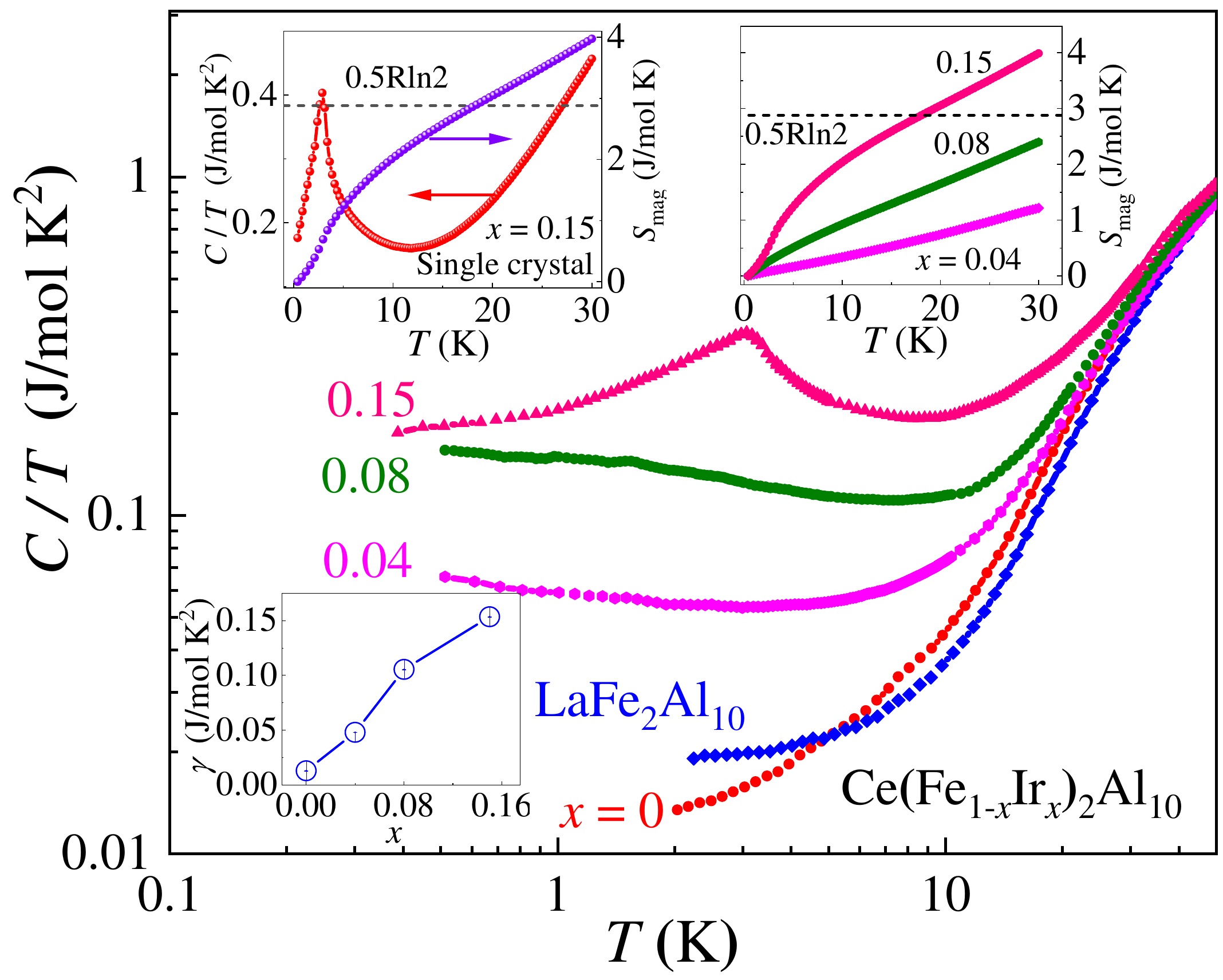}
		\caption{Temperature dependence of $C/T$ for polycrystalline samples of  Ce(Fe$_{1-x}$Ir$_x$)$_2$Al$_{10}$ ($x = 0$, 0.04, 0.08, 0.15) and  LaFe$_2$Al$_{10}$, on a log–log plot. Upper right inset: Magnetic entropy $S_{\mathrm{mag}}$ as a function of temperature for the polycrystalline samples. Lower inset: Variation of $\gamma$ [calculated in the temperature range $\left(7 \lesssim T \lesssim 20~\mathrm{K} \right)$] as a function of Ir concentration $x$ for the polycrystalline samples. Upper left inset: $C/T$ versus $T$ (left axis) and $S_{\mathrm{mag}}$ versus $T$ (right axis) for a single crystal sample of $x = 0.15$.}
		\label{SpecificHeat}
	\end{figure}

	The double logarithmic plots of $C/T$ versus $T$ for Ce(Fe$_{1-x}$Ir$_x$)$_2$Al$_{10}$ ($x = 0$, 0.04, 0.08, 0.15) and LaFe$_2$Al$_{10}$, are shown in  Fig.~\ref{SpecificHeat}. For $x=0$, $C/T$ decreases with decreasing temperature and becomes smaller that its La counterpart, supporting the existence of a pseudo-gap at $E_{\mathrm{F}}$. For $x = 0.04$ and 0.08, the low-temperature $C/T$ is considerably enhanced without any sign of magnetic ordering, reminiscent of heavy-fermion-like behavior. For $x = 0.15$, however, a $\lambda$-type anomaly is observed at $T \sim 3$~K. This is the same temperature at which the $\chi(T)$ data exhibit a peak (see Section~\ref{MagSusc}), indicating the onset of long-range AFM ordering for $x = 0.15$. Here it is important to note that no magnetic ordering was seen down to 2~K in Ce(Fe$_{0.8}$Rh$_{0.2}$)$_2$Al$_{10}$, despite the enhancement of the localized nature of the $4f$ electron state~\cite{PhysRevB.92.235154}. 
	
	At temperatures 7-20~K $C/T$ varies linearly with $T^2$. Fitting to the data with the form $C={\gamma}T+\beta T^{3}$, the values of $\gamma$ were estimated as function of $x$ as shown in Fig.~\ref{SpecificHeat}. For $x = 0$, $\gamma$ is almost zero, suggesting a very small or zero density of states at the Fermi level. However, for $x = 0.04$, $\gamma$ is dramatically enhanced and reaches 0.154(3)~J/mole K$^2$ for $x = 0.15$. In other $T$-site substituted systems e.g. Ce(Ru$_{1-x}$Rh$_x$)$_2$Al$_{10}$~\cite{PhysRevB.90.165124} or Ce(Os$_{1-x}$Ir$_x$)$_2$Al$_{10}$~\cite{PhysRevB.89.094404}, the $\gamma$ value increases rapidly with $x$, which was attributed to a rapid collapse of the spin gap and the appearance of conduction electrons with high effective mass at the Fermi level. We therefore anticipate that the enhancement of $\gamma$ in the specific heat of Ce(Fe$_{1-x}$Ir$_x$)$_2$Al$_{10}$ is associated with a change in the gap structure (especially the charge or transport gap in $x=$ 0.04) and the existence of a finite density of states close to the Fermi level. Support for this scenario comes from the behavior of the resistivity shown in Fig.~\ref{resistivity}(b). In addition, it is to be noted that our results for electron doping differ from those of isoelectronic substitution in Ce(Fe$_{1-x}$Ru$_x$)$_2$Al$_{10}$, where the Kondo semiconducting ground state with the gap is maintained in all the solid solutions~\cite{PhysRevB.87.224415,PhysRevB.84.165125}. This suggests that the observed behavior in the Ir-substituted CeFe$_2$Al$_{10}$ is essentially due to an excess of electrons rather than any lattice imperfections or disorder, because Ru substitution produces the same amount of disorder but does not produce any change in the Kondo semiconducting ground state. As such, it seems that the electron doping achieved by substituting  Fe with Ir (or Rh), weakens the $c\mhyphen f$  hybridization and destroys the Kondo semiconducting character and the spin/charge gapped state at low temperatures. The observation of a long-range ordered magnetic ground state with Ir substitution rather than the paramagnetic ground state seen with Rh substitution could be attributed to the $b$ axis behavior and also strength of spin-orbit effects. 
	
	The contribution of $4f$ electrons to the heat capacity was estimated as $C_{4f}$  = $C$[Ce(Fe$_{1-x}$Ir$_{x}$)$_2$Al$_{10}$] - $C$[LaFe$_2$Al$_{10}$]. The integration of $C_{4f}/T$ with respect to $T$ gives the magnetic entropy $S_{\mathrm{mag}}$, as shown in the upper right inset of Fig.~\ref{SpecificHeat}. For $x = 0.15$, $S_{\mathrm{mag}}$ is $0.11\mathrm{R}\ln 2$ at $T_{\mathrm{N}}=3.1$~K and $0.87\mathrm{R}\ln 2$ at 30~K. The reduced value of $S_{\mathrm{mag}}$ indicates the presence of strong Kondo screening of the $4f$ moments by the conduction electrons even in the magnetically ordered state. Another possible source for a reduced entropy could be the presence of short-range magnetic fluctuations that exist at temperatures well above $T_{\mathrm{N}}$. A reduced value of $S_{\mathrm{mag}} = 0.3\mathrm{R}\ln 2$ at $T_{\mathrm{N}}$ has also been observed in CeOs$_2$Al$_{10}$ \cite{PhysRevB.81.214401}. 
	
	The single crystal sample of Ce(Fe$_{0.85}$Ir$_{0.15}$)$_2$Al$_{10}$ displays a sharp $\lambda$ type anomaly in $C\left(T\right)$ at 3~K, as shown in the upper left inset of Fig.~\ref{SpecificHeat}, which is consistent with the data for the polycrystalline sample of $x = 0.15$. The $S_{\mathrm{mag}}$($T$) for the single crystal agrees well with that of the polycrystalline sample with $x=0.15$. 
	
	\subsection{Magnetic Susceptibility}
	\label{MagSusc}
	
	\begin{figure}
		\includegraphics[width=8.5cm, keepaspectratio]{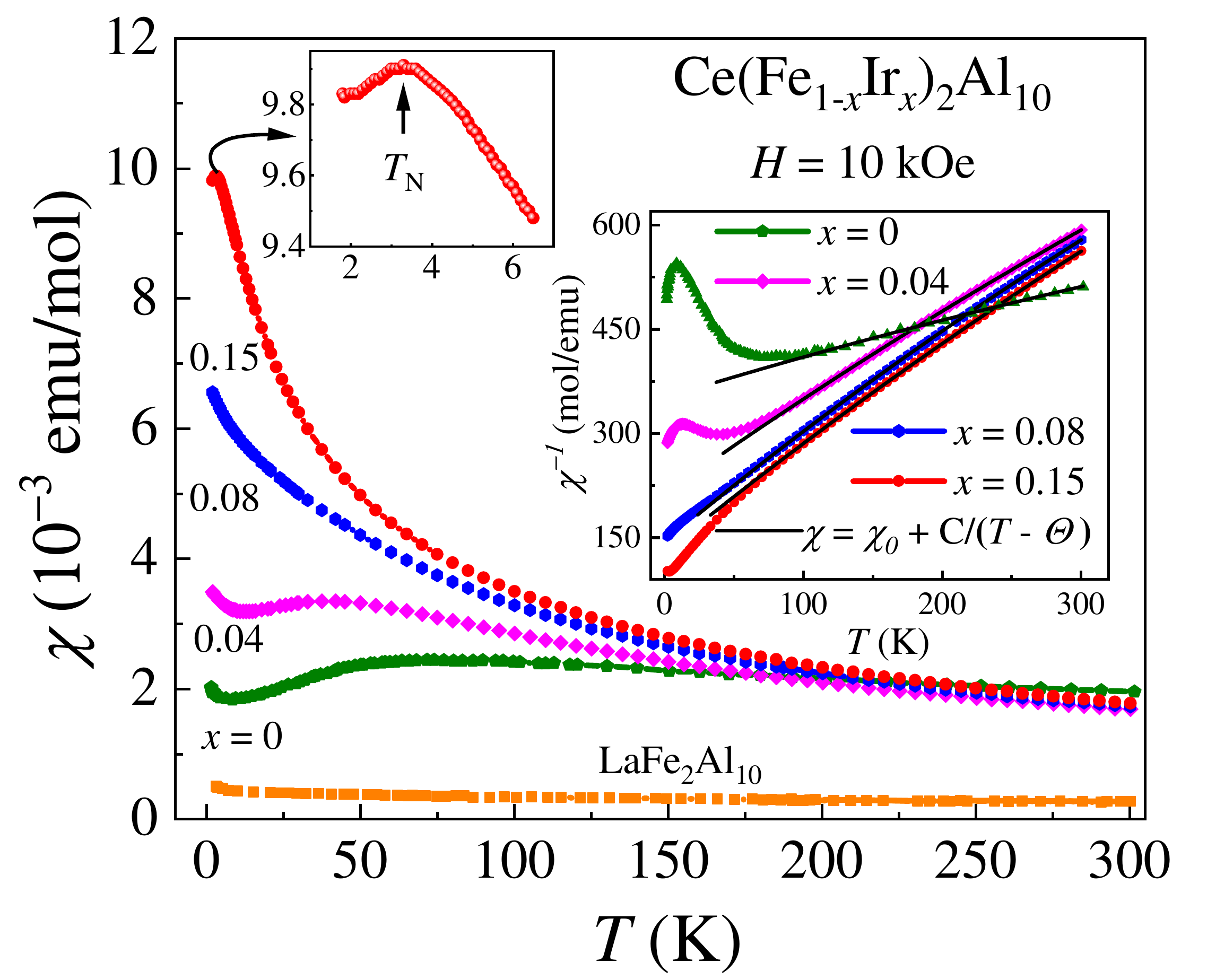}
		\caption{Temperature dependence of magnetic susceptibility $\chi$($T$) for polycrystalline samples of Ce(Fe$_{1-x}$Ir$_x$)$_2$Al$_{10}$ ($x = 0$, 0.04, 0.08, 0.15) and LaFe$_2$Al$_{10}$ measured at $H = 10$~kOe. Upper left inset shows the peak in $\chi\left(T\right)$ for $x = 0.15$. Lower right inset shows $\chi^{-1}$ vs $T$ for $x = 0$, 0.04, 0.08, 0.15. The solid lines are fits to the data with the modified Curie-Weiss behavior including a temperature independent term $\chi_0$.}
		\label{MSusc1}
	\end{figure}

	\begin{figure}
		\includegraphics[width=8.5cm, keepaspectratio]{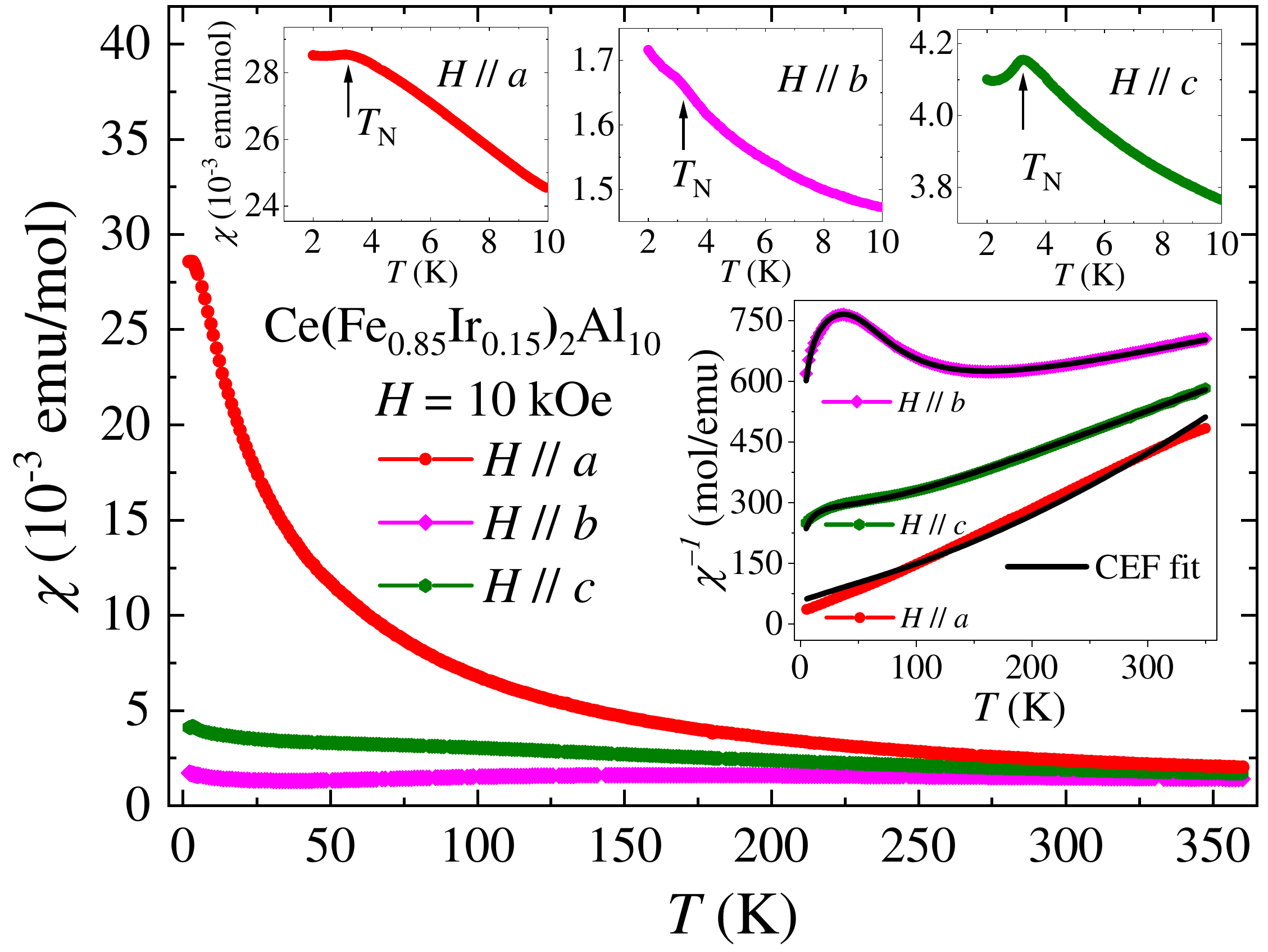}
		\caption{ Temperature dependence of magnetic susceptibility $\chi$($T$) of single crystal Ce(Fe$_{0.85}$Ir$_{0.15}$)$_2$Al$_{10}$ in a field of 10~kOe applied along the three principal axes. Upper insets: $\chi$ vs $T$ at low temperature. Lower inset: $\chi^{-1}$ vs $T$ along the three principal axes. The solid lines are fits to the data with the crystal electric field (CEF) model discussed in the text.}
		\label{MSusc2}
	\end{figure}

	Figure~\ref{MSusc1} shows the temperature dependence of the magnetic susceptibility $\chi(T) = M(T)/H$ [$H = 10$~kOe] for polycrystalline samples of Ce(Fe$_{1-x}$Ir$_x$)$_2$Al$_{10}$ ($x = 0$, 0.04, 0.08, 0.15) and LaFe$_2$Al$_{10}$. At $T\> 100$~K, $\chi^{-1}$ vs $T$ data shows a Curie-Weiss (CW) behavior as shown in the lower inset of Fig.~\ref{MSusc1}. The broad peak in $\chi(T)$ near 75~K for $x = 0$ moves to 40~K for $x = 0.04$ and then disappears for $x = 0.08$. A similar trend in $\chi(T)$ with electron doping was also seen for Rh-doped CeFe$_2$Al$_{10}$ and was attributed to a localization of $4f$ electrons as a consequence of an increase in the Fermi energy by electron doping~\cite{PhysRevB.92.235154}. However, no signature of magnetic ordering could be seen down to 2~K for up to 20\% Rh doping. On the other hand, $\chi(T)$ of Ce(Fe$_{1-x}$Ir$_x$)$_2$Al$_{10}$ for $x=0.15$ shows an anomaly at $T = 3.1(2)$~K (an enlarged view is shown in the upper inset of Fig.~\ref{MSusc1}), indicating a phase transition to an AFM ordered state. A least squares fit to the $\chi^{-1}$ vs $T$ data above 100~K with the Curie-Weiss law including a temperature independent term (lower inset of Fig.~\ref{MSusc1}), yields an effective magnetic moment $\mu_{\mathrm{eff}} = 2.78(2)$, 2.23(1), 2.10(1) and 2.12(1) $\mu_{\mathrm{B}}$/Ce, and paramagnetic Weiss temperatures $\Theta$ of --439(11), --140(2), --81.9(1), and --77.2(2)~K for the samples with $x = 0$, 0.04, 0.08, and 0.15, respectively. The value of $\mu_{\mathrm{eff}}$ is slightly larger than that of a free Ce$^{3+}$ ion (2.54~$\mu_{\mathrm{B}}$) for $x = 0$, while it is slightly smaller than expected for the Ir-doped alloys. This discrepancy may be due to crystal-field effects in $\chi(T)$. A large and negative value of $\Theta$ is a common feature in Ce-based compounds with strong $c\mhyphen f$ hybridization~\cite{PhysRevB.75.024432}. The decrease of $\left|\Theta \right|$ from 439~K for $x = 0$ to 77~K for $x = 0.15$ also suggests a decrease in $T_{\mathrm{K}}$ because the value of $T_{\mathrm{K}}$ for the overall CEF levels is proportional to $\left|\Theta \right|$~ \cite{doi:10.1080/00018738400101681}. 
	
	Magnetization isotherms for the polycrystalline samples of Ce(Fe$_{1-x}$Ir$_x$)$_2$Al$_{10}$ ($x = 0.04$, 0.08, 0.15), measured at various temperatures (see Supplemental Material~\cite{SM}), increase linearly with field. The magnetization at fixed temperature and field increases with $x$. However, the magnetization for $x = 0.15$ (0.12 $\mu_{\mathrm{B}}$/Ce at $H = 70$~kOe) is significantly smaller than the theoretical saturation magnetization of $gJ$ = 2.14~$\mu_{\mathrm{B}}$ for a free ion of Ce$^{3+}$. The magnetization data thus hint at a weak AFM staggered moment in the magnetically ordered state which is most likely a consequence of a Kondo effect. Further evidence of the reduced moment ordering in $x=0.15$ comes from the muon spin rotation study presented in Section~\ref{PNDMuSR}.
	
	Now, we discuss the anisotropy of $\chi$($T$) of the single crystalline sample with $x=0.15$. As shown in Fig.~\ref{MSusc2}, the observed susceptibility is highly anisotropic, i.e., $\chi_a \gg \chi_c > \chi_b$ over the whole measured temperature range. The easy-magnetization along the $a$ axis is a common characteristic among the Ce$T_2$Al$_{10}$ family. The anisotropy in $\chi\left(T\right)$ mainly arises from the CEF effects as seen in CeRu$_2$Al$_{10}$ and CeOs$_2$Al$_{10}$~\cite{PhysRevB.81.214401,doi:10.1143/JPSJ.79.083701}. As presented in the supplementary material~\cite{SM}, the magnetization $M_a$ with $H \parallel a$ is strongly enhanced over $M_b$ and  $M_c$ below $T_{\mathrm{N}}$~\cite{SM}. The magnetic moment per Ce ion along the easy $a$ axis reaches only 0.25~$\mu_{\mathrm{B}}$ at 50~kOe, which is still far from the $2.14~\mu_{\mathrm{B}}$/Ce expected for a free Ce$^{3+}$ ion. $M$ vs $H$ increases linearly up to 50~kOe without showing a spin-flip transition. The upper insets of Fig.~\ref{MSusc2} show $\chi(T)$ along the three principal axes at low temperature. For $H \parallel c$ there is a clear maximum at $T_{\mathrm{N}}= 3.1(2)$~K while for $H \parallel a$ and $b$, a more diffuse feature is centered at $T_{\mathrm{N}}$. This is in agreement with other compounds of this family, where pronounced peaks in $\chi(T)$ are found for $H \parallel c$ and $H \parallel a$~\cite{PhysRevB.89.094404,PhysRevB.90.165124}.
	
	At higher temperatures $\chi_b(T)$ and $\chi_c(T)$ exhibit a broad maximum at $T \sim 150$~K and a shoulder at $T \sim 75$~K, respectively, well above $T_{\mathrm{N}}$, that can be attributed to CEF effects. The point symmetry of the Ce ion on the $4c$ site is orthorhombic $C_{2v}$ and hence the CEF Hamiltonian, with the quantization axis along the $b$~axis, is given by the following expression
	
	\begin{equation}
		H_{\mathrm{CF}}=B_{2}^{0}O_{2}^{0}+B_{2}^{2}O_{2}^{2}+B_{4}^{0}O_{4}^{0}+B_{4}^{2}O_{4}^{2}+B_{4}^{4}O_{4}^{4},
	\end{equation}
	where $B_{n}^{m}$ are the CEF parameters and $O_{n}^{m}$ are the Stevens operators~\cite {Stevens}. Analysis of single crystal susceptibility data to determine the CEF parameters was carried out using the software available in the Mantid program~\cite{MantidCEF}. Fits to the data at 5 - 350~K are shown by the solid black lines on the $\chi^{-1}$ vs $T$ plots in the lower inset of Fig.~\ref{MSusc2}. The final set of CEF parameters are (in K) $B_{2}^{0}=0.9607(4)$, $B_{2}^{2}=-1.60(3)$, $B_{4}^{0}=-0.0437(2)$, $B_{4}^{2}=0.2101(4)$, and $B_{4}^{4}=0.3930(1)$. In addition, the CEF splitting energies of the first ($\Delta_1$) and second ($\Delta_2$) excited states are 240~K (20~meV) and 422~K (38~meV), respectively. It is worth noting that the overall splitting is rather small compared to that of CeFe$_2$Al$_{10}$, discussed in the next section.

	Here, it is important to compare the results of the Ce(Fe$_{1-x}$Ir$_x$)$_2$Al$_{10}$ and Ce(Fe$_{1-x}$Rh$_x$)$_2$Al$_{10}$ systems in order to discuss the origin of the magnetic ordering. The substitution of either Rh or Ir for Fe in CeFe$_2$Al$_{10}$ leads to the addition of electrons. The substitution of Rh up to 20\% does not induce magnetic order. However, in the case of Ce(Fe$_{0.85}$Ir$_{0.15}$)$_2$Al$_{10}$, this results in a magnetically ordered ground state. Here we suggest that the $4d$-electron doping weakens the Kondo semiconducting character of CeFe$_2$Al$_{10}$ but not sufficiently to lead to the onset of magnetic order.  We propose that two contributions lead to the magnetic ground state in Ir-doped CeFe$_2$Al$_{10}$. One is the increase in the density of states at the Fermi level due to the doped electrons, and the other is the contraction along the $b$ axis, as discussed in section~\ref{structure}. Here it is important to note that in Ce$T_2$Al$_{10}$ ($T = $~Ru, Os), there is a gap related to the charge- and spin-density waves along the $b$ axis, in addition to a hybridization gap along all three axes~\cite{PhysRevB.91.241120,PhysRevLett.106.056404}. Moreover, it is also found that the application of uniaxial pressure $P\parallel b$ strongly enhanced the $T_{\mathrm{N}}$ of Ce$T_2$Al$_{10}$ ($T = $~Ru, Os). In Ce(Fe$_{0.85}$Ir$_{0.15}$)$_2$Al$_{10}$ we have observed contraction of $b$ axis, which may be taken as an effective uniaxial pressure in the system and hence gives magnetic ordering compared to Rh doping.
	Our results further suggest that the hybridization gap is not mandatory for the AFM ordering observed in the $x = 0.15$ sample, since the low-temperature Kondo semiconducting increase in $\rho$ disappeared and shows a metallic ground state along with a long-range AFM ordering in $C\left(T\right)$ and $\chi\left(T\right)$.
	
	Further, it is to be noted that spin-orbit coupling (SOC) (which varies as $Z^4$, where $Z$ is atomic number) is stronger for Ir than for Rh and hence the SOC may also be playing an important role in the observed differences.
	
	\subsection{INELASTIC NEUTRON SCATTERING}
	\label{INS}
\begin{figure}
		\includegraphics[width=7.5cm, keepaspectratio]{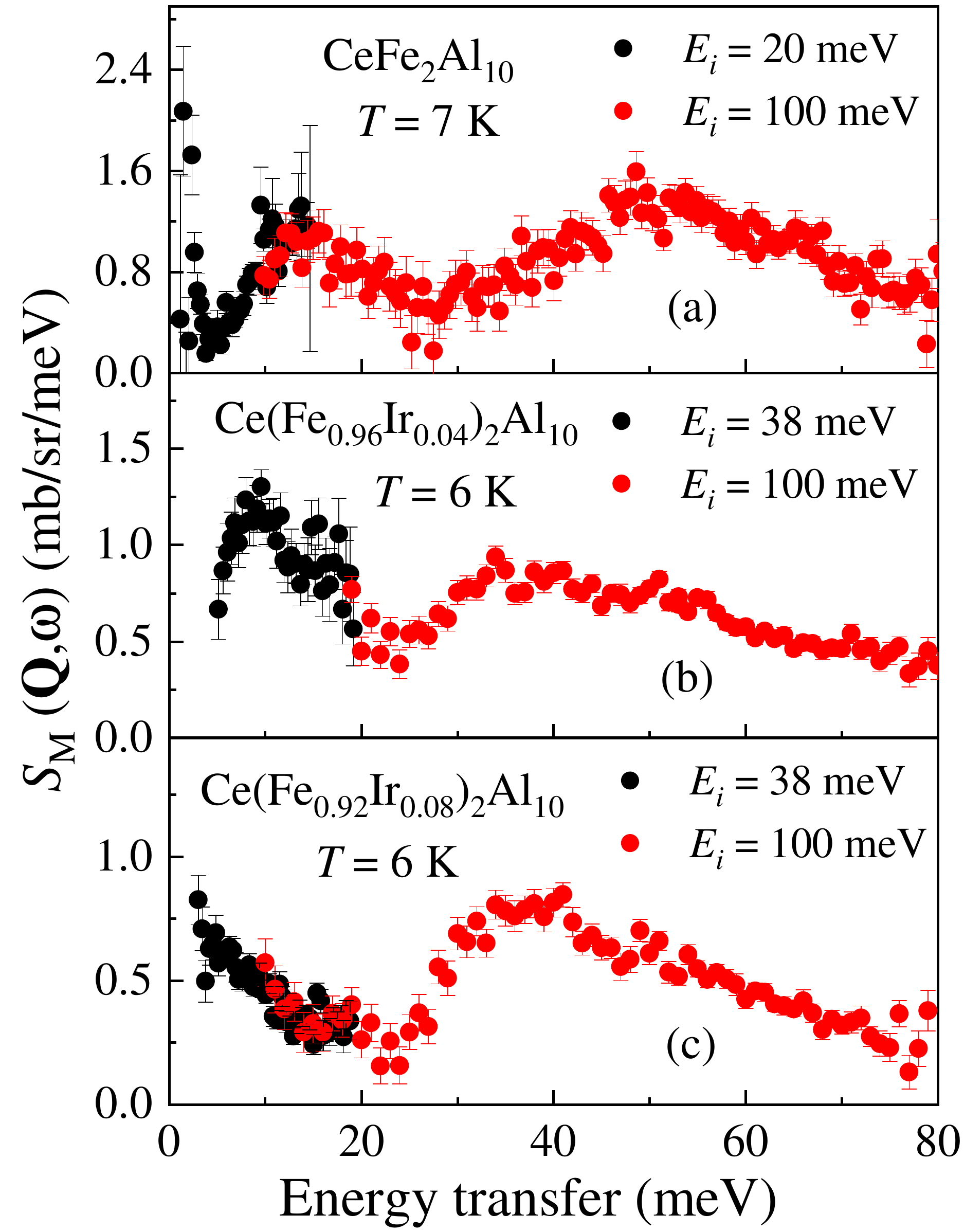}
		\caption{Inelastic neutron scattering data at 7~K after subtracting the phonon contribution for (a) CeFe$_2$Al$_{10}$ (data are from Ref.~\onlinecite{PhysRevB.87.224415} and are given for comparison), (b) Ce(Fe$_{0.96}$Ir$_{0.04}$)$_2$Al$_{10}$, and (c) Ce(Fe$_{0.92}$Ir$_{0.08}$)$_2$Al$_{10}$ at 6~K.}
		\label{Neutrons}
	\end{figure}

	We have performed inelastic neutron scattering (INS) measurements on polycrystalline samples of Ce(Fe$_{1-x}$Ir$_{x}$)$_2$Al$_{10}$ with $x = 0.04$  and 0.08, in order to investigate whether the spin gap has disappeared and to determine the crystal field excitations. We also measured LaFe$_2$Al$_{10}$ in order to estimate the phonon scattering. The estimated magnetic scattering after subtracting the phonon scattering is shown in Figs.~\ref{Neutrons}(b-c). For comparison, in Fig.~\ref{Neutrons}(a) we also present the results for CeFe$_2$Al$_{10}$  taken from Ref.~\onlinecite{PhysRevB.87.224415}. Fig.~\ref{Neutrons}(b) shows that for $x = 0.04$ the spin gap is still present, but the gap value is reduced to 8~meV from 12.5~meV for $x = 0$. On the other hand  for $x = 0.08$ the spin gap has closed completely and the low-energy response transforms into broad quasi-elastic scattering as shown in Fig.~\ref{Neutrons}(c). For $x = 0.04$, a spin gap of 8~meV exists in the INS spectrum, but there is no charge gap in the resistivity down to 2~K (see Fig.~\ref{resistivity}). A very similar behavior has been observed in Ce(Fe$_{1-x}$Rh$_{x}$)$_2$Al$_{10}$ with $x = 0.05$, which has a spin gap of 9(2)~meV in its INS spectrum~\cite{PhysRevB.102.024438}, but has no charge gap in the resistivity \cite{PhysRevB.92.235154}. We attribute this behavior to either the development of an in-gap density of states or a shift of the Fermi level position from the middle of the lower and upper hybridized band (i.e. $E_{\mathrm{F}}$ positioned in the the gap) to the bottom of the upper hybridized band. The high-energy peak observed near 55~meV in $x = 0$ has moved to lower energy, 40 and 37~meV, for $x =$ 0.04 and 0.08, respectively. This lowering in the position of the high-energy peak is due to a reduction in the $c \mhyphen f$ hybridization. It is surprising that even though the susceptibility of $x = 0.08$ reveals Curie-Weiss behavior of trivalent Ce ions (see inset of Fig.~\ref{MSusc1}), the CEF excitation near 37~meV is very broad. Now let us compare the CEF excitations observed in $x = 0.08$ with that observed in CeRu$_2$Al$_{10}$ and CeOs$_2$Al$_{10}$~\cite{AdrojaUnpublished}. In CeRu$_2$Al$_{10}$, there are two well-defined CEF excitations at 30 and 45~meV at 5~K. At 44~K (i.e. above $T_{\mathrm{N}}=27$~K), the two clearly resolved CEF excitations are transformed into one broad excitation near 30~meV with a tail up to 80~meV (a second possible CEF excitation is seen near 45~meV in the tail).  In CeOs$_2$Al$_{10}$ the CEF excitations are broad and not well-resolved  both at 5 and 65~K ($T_{\mathrm{N}}=28.9$~K). The first CEF excitation is at 36~meV with a tail going up to 80~meV with the possibility of a second CEF excitation near 60~meV. The high energy CEF response seen here in Ce(Fe$_{1-x}$Ir$_{x}$)$_2$Al$_{10}$ with $x = 0.08$ [Fig.~\ref{Neutrons}(c)] is very similar to that observed for CeOs$_2$Al$_{10}$.
	
	\subsection{NEUTRON DIFFRACTION AND MUON SPIN RELAXATION}
	\label{PNDMuSR}
	
	\begin{figure}
		\includegraphics[width=7.5cm, keepaspectratio]{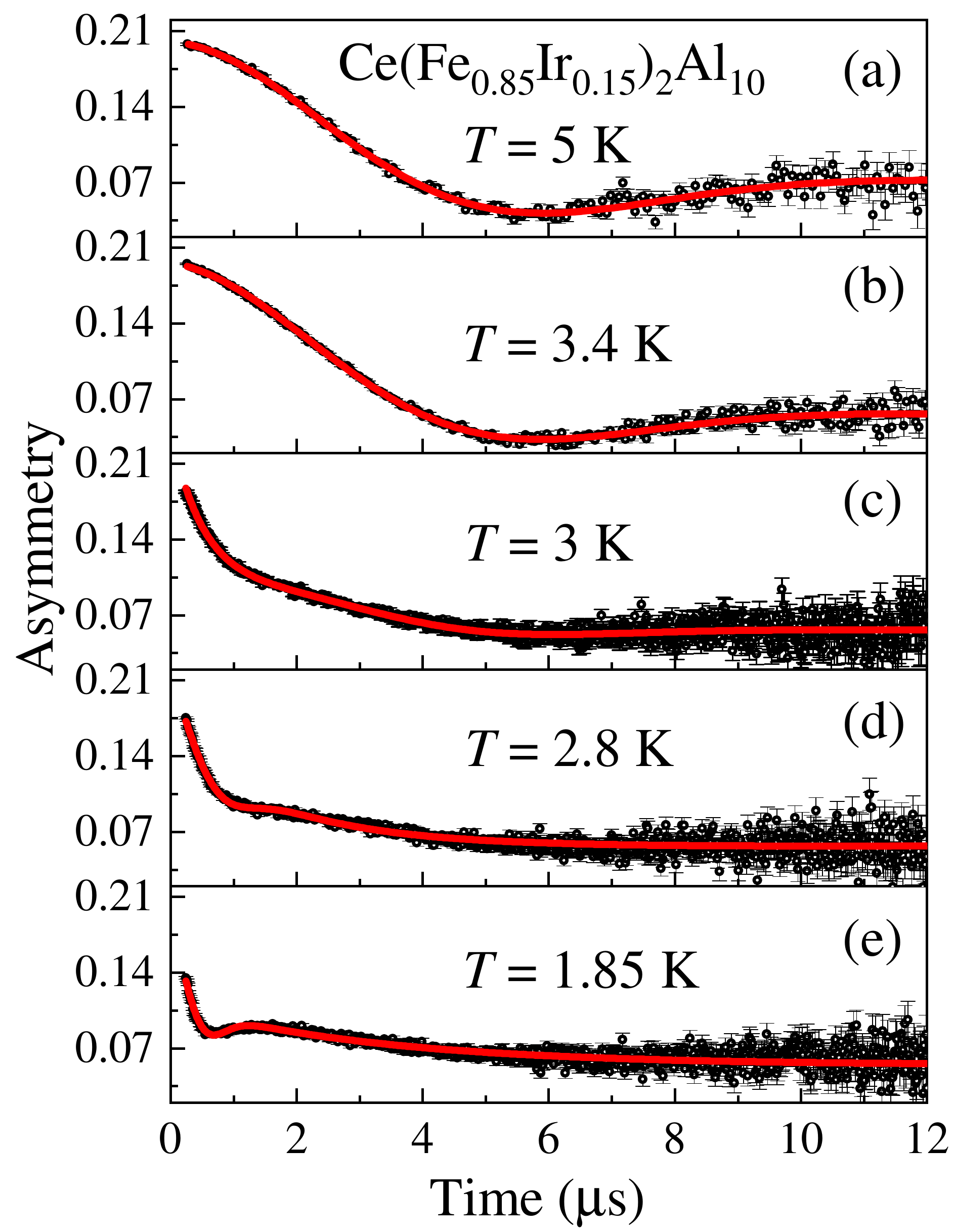}
		\caption{ Zero-field $\mu$SR spectra for polycrystalline Ce(Fe$_{1-x}$Ir$_{x}$)$_2$Al$_{10}$ ($x = 0.15$) at various temperatures. The solid lines are least-squares fit to the data as described in the text.}
		\label{MUSR1}
	\end{figure}
	
	\begin{figure}
		\includegraphics[width=8.5cm, keepaspectratio]{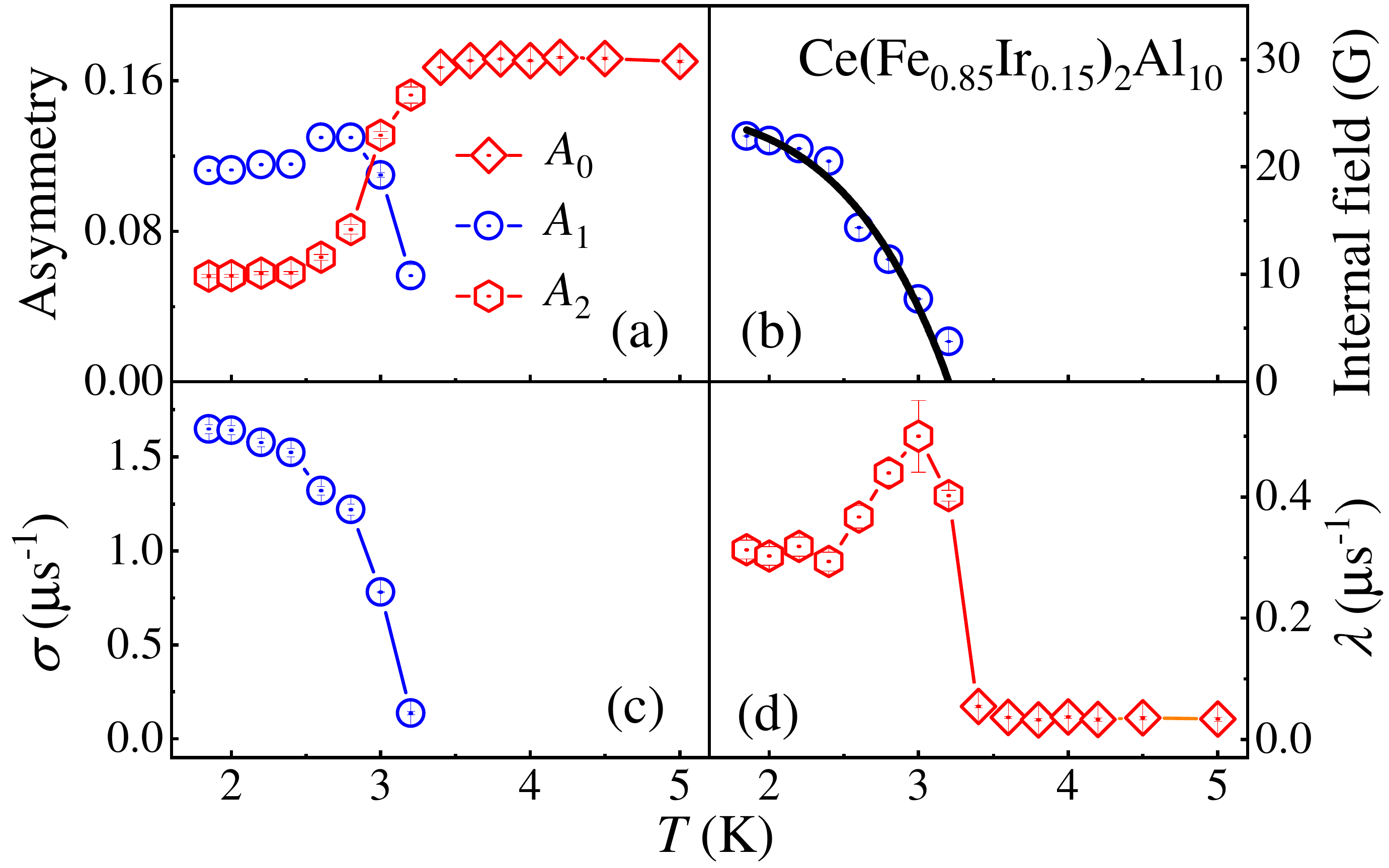}
		\caption{Temperature dependence of (a) the initial asymmetry, (b) the internal field at the muon stopping site, obtained by zero-field $\mu$SR experiment on Ce(Fe$_{0.85}$Ir$_{0.15}$)$_2$Al$_{10}$. The solid line represents fits using Eq.~\ref{Hint}. (c) Temperature dependence of the distribution width of the local Gaussian field and (d) the depolarization rate.}
		\label{MUSR2}
	\end{figure}
	
	In order to understand the magnetic interactions and magnetic structure at the microscopic level, we have performed neutron diffraction measurements on a polycrystalline sample of Ce(Fe$_{1-x}$Ir$_{x}$)$_2$Al$_{10}$ with $x = 0.15$ using the WISH diffractometer (see Supplemental Material~\cite{SM}). 
	No clear sign of magnetic Bragg peaks could be observed in the diffraction data collected below $T_{\mathrm{N}}$, which indicates that the magnetic moment in the ordered-state is very small. Our simulation of the magnetic scattering, using models reported previously for undoped and carrier-doped Ce$T_2$Al$_{10}$ ($T = $~Ru and Os) systems~\cite{PhysRevB.82.100405,doi:10.1143/JPSJ.80.073701}, allowed us to set an upper limit for the ordered moment of $\sim 0.07(1)~\mu_{\mathrm{B}}$/Ce. 
	
	Given the very small ordered moment for $x = 0.15$, we performed a positive muon spin relaxation measurement. The muon spin relaxation technique is a very sensitive local probe for characterizing static and dynamic magnetism and has been extensively used to trace the onset of AFM order in Ce$T_2$Al$_{10}$ materials~\cite{PhysRevB.82.100405, PhysRevB.82.104405}. The zero-field-$\mu$SR asymmetry spectra for $x = 0.15$ collected at several temperatures ranging from 1.8 to 5~K, are shown in Figs.~\ref{MUSR1}(a) - \ref{MUSR1}(d). As shown in Figs.~\ref{MUSR1}(a) and~\ref{MUSR1}(b) at 3.4-5~K, the ZF-$\mu$SR spectra show a Kubo-Toyabe (KT) type behavior, which originates from a static internal field with a Gaussian distribution of nuclear dipole moments~\cite{PhysRevB.20.850}. In the paramagnetic regime the $\mu$SR spectra can be described by:
	
	\begin{equation}
		A(t)=A_0 G_{\mathrm{KT}}(t) \exp \left(-\lambda t\right) + A_{\mathrm{bg}},
		\label{MUSRHighT1}
	\end{equation}
	
	\noindent where, 
	
	\begin{equation}
		G_{\mathrm{KT}}(t)=\frac{1}{3}+\frac{2}{3}\left(1-\Delta^{2} t^{2}\right) \exp \left(-\frac{\Delta^{2} t^{2}}{2}\right),
		\label{MUSRHighT2}
	\end{equation}
	
	\noindent is the Kubo-Toyabe function, $A_0$ is the initial asymmetry at $t=0, \Delta / \gamma_{\mu}$ represents the distribution width of the local Gaussian fields, $\gamma_{\mu} / 2 \pi = 135.5 \mathrm{MHz} / \mathrm{T}$ is the muon gyromagnetic ratio, $\lambda$ is the depolarization rate caused by fluctuating electronic spins, and $A_{\mathrm{bg}}$ is a constant background. Best fits to the spectra using Eqs. (\ref{MUSRHighT1}) and (\ref{MUSRHighT2}), shown by the solid line in Fig.~\ref{MUSR1}(a), reveal a nearly $T$ independent $\Delta$ equal to $0.295(1)~\mu\mathrm{s}^{-1}$ above $T_{\mathrm{N}}$. The background component arising from muons stopping in the silver sample holder, was estimated to be $A_{\mathrm{bg}}=0.027$ from the 3.4 to 5~K data. 
	
	The Larmor precession of the muon spin around the internal fields of the magnetically ordered system gives rise to an oscillatory muon polarization as a function of time $t$. Our $\mu$SR spectra exhibit oscillations below 3.2~K, confirming the long-range magnetic ordering of the Ce moments. Figures~\ref{MUSR1}(c) - ~\ref{MUSR1}(d) show these spectra can be well-fitted by 
	\begin{equation}
		G_{z}(t)=  A_{1} \cos \left(\omega t+\phi\right) \exp\left(-\sigma^2t^{2} / 2\right)+ A_{2} \exp\left(-\lambda t\right) + A_{\mathrm{bg}},
		\label{Gzt}
	\end{equation} 
	
	\noindent where $A_1$, $A_2$, $\lambda$, $\omega \left(= \gamma_{\mu}B_\mathrm{{int}}\right)$, and $\phi$, are the associated asymmetries, exponential depolarization rate, muon spin rotation frequency ($B_{\mathrm{int}}$ is the internal field at the muon site), and the initial phase, respectively. The first term of Eq. (\ref{Gzt}) represents the transverse components of the internal fields seen by the muons along which they precess, while the second term represents the longitudinal component. A different value of $A_{\mathrm{bg}}=0.054$ was observed in temperature range 1.85 to 3.2~K. This indicates that below 3.4~K an additional oscillatory component with an initial asymmetry $\sim$~0.027 may be present, cf. CeRu$_2$Al$_{10}$ \cite{PhysRevB.82.100405}. However, due the very low values of the frequency and relaxation rate of this component, their values were set to zero, and the component folded into $A_{\mathrm{bg}}$ in order to allow the fit to converge. The fitting parameters as a function of temperature, determined from the best fits, are shown in Figs.~\ref{MUSR2}(a)~–~\ref{MUSR2}(d). Figure~\ref{MUSR2}(b), shows the internal field (muon precession frequency) at the muon site as a function of temperature. This shows that the internal field (or frequency) appears just below 3.2~K, indicating the onset of bulk long-range magnetic order, which agrees with the specific heat and magnetic susceptibility data. The value of the internal field is 22.4~G at the base temperature (i.e. 1.85~K). This is smaller than the maximum field seen in CeRu$_2$Al$_{10}$ (120~G)~\cite{PhysRevB.82.100405} and CeOs$_2$Al$_{10}$~(50~G)~ with the ordered moment along the $c$~axis~\cite{PhysRevB.82.104405}. For Ce(Fe$_{0.85}$Ir$_{0.15}$)$_2$Al$_{10}$, we assumed that the ordered moment lies along the easy axis of the CEF (i.e. $a$~axis). A rough estimate of the ordered moment from the observed internal field at the muon stopping site, made using the results of the dipolar field calculation from Ref.~\onlinecite{PhysRevB.88.115206}, gives 0.42~$\mu_{\mathrm{B}}/\mathrm{Ce}\times(22.4/142) = 0.066~\mu_{\mathrm{B}}/\mathrm{Ce}$. This is in agreement with the absence of visible magnetic Bragg peaks in the WISH data. Below 3~K, the depolarization rate also increases as shown in Fig.~\ref{MUSR2}(d). In principle, this could originate from various phenomena related to a change in the distribution of internal fields. The relaxation rate $\lambda$ associated with the local field fluctuation rates sensed by the muon spin exhibits a sharp peak at $T = 3$~K. Notably, such behavior in $\lambda$ is expected across a magnetic phase transition.

	We examined the temperature dependence of the internal fields, which appear at temperatures below 3.2~K and signify the onset of long-range magnetic order, in order to determine the nature of the magnetic interactions. The temperature dependence of the internal field was fitted using
	\begin{equation}
		B_{\mathrm{int}}(T)=B_{0}\left[1-\left(\frac{T}{T_{N}}\right)^{\alpha}\right]^{\beta},
		\label{Hint}
	\end{equation}
	\noindent The fit is shown by the solid curve in Fig.~\ref{MUSR2}(b). The fitting parameters are $B_{0}=25(4) $~G, $\beta = 0.97(5)$,  $\alpha = 4(3)$, and $T_{\mathrm{N}} = 3.2(1)$~K. A fit with $\beta \sim 0.97$ suggests the magnetic interactions in the electron-doped system are not purely of a mean-field type for which $\beta$ is expected to be 0.5. In fact, $\alpha > 1$ indicates complex magnetic interactions within this system~\cite{D.I.K,S.B}.

	\begin{figure}
		\includegraphics[width=8.5cm, keepaspectratio]{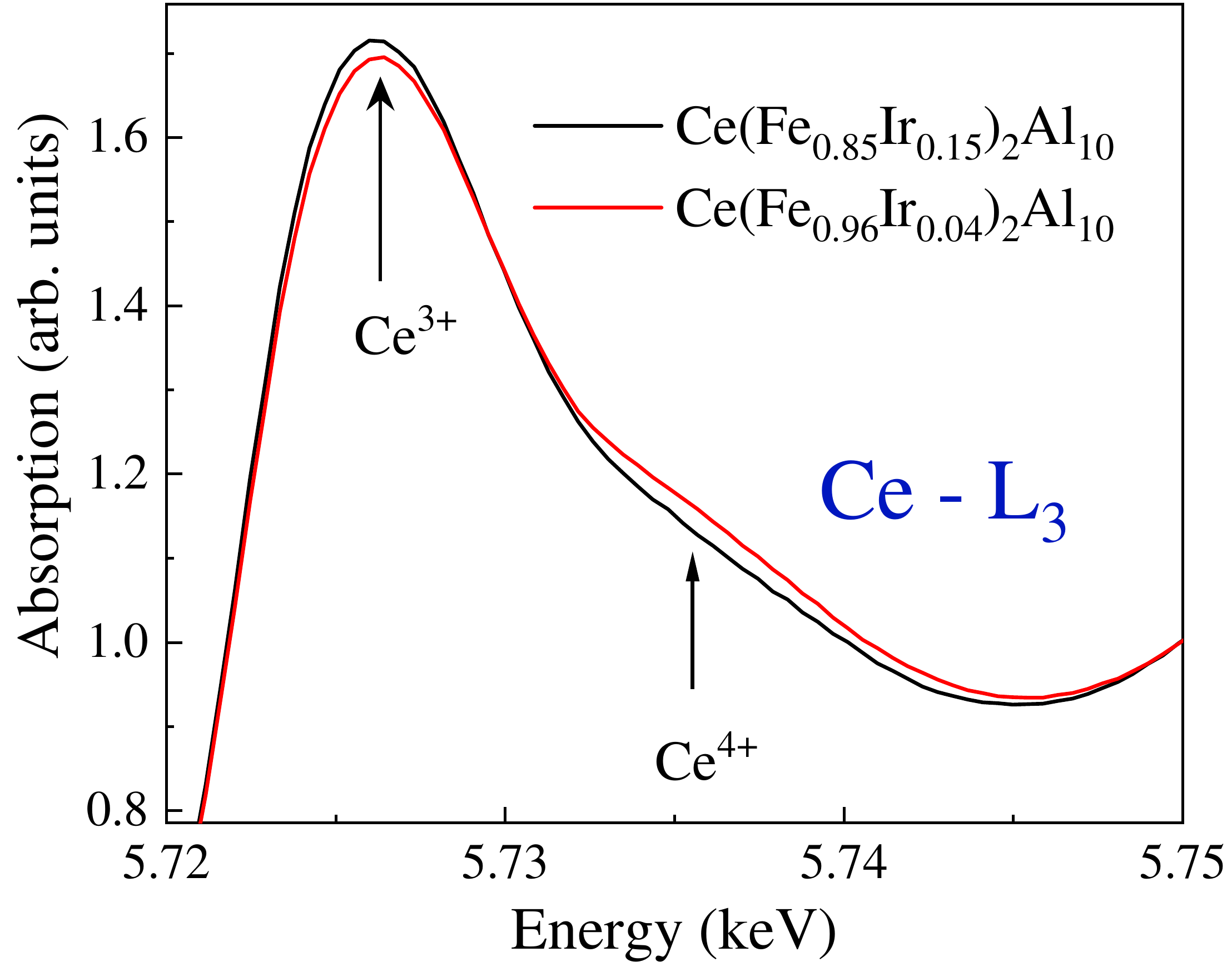}
		\caption{Ce L$_3$ x-ray absorption edge spectra of Ce(Fe$_{1-x}$Ir$_{x}$)$_2$Al$_{10}$ for $x = 0.04$ and 0.15 at 7~K.}
		\label{XANES}
	\end{figure}

	\subsection{Ce L$_3$-edge x-ray absorption spectroscopy}
	\label{CeLedge}
	We performed Ce L$_3$-edge x-ray absorption near-edge spectroscopy (XANES) at 7~K on polycrystalline samples with $x = 0.04$ and 0.15, in order to obtain more information about the valence state of the Ce ions as a function of Ir-doping. The spectrum is sensitive to electronic transitions from the core level to the higher unfilled or half-filled orbitals of the absorbing atom. XANES is therefore uniquely placed to measure valence states. As shown in Fig.~\ref{XANES}, the spectra for both $x = 0.04$ and 0.15 have a prominent peak at about 5728~eV, which corresponds to a signal from a bulk Ce$^{3+}$ configuration. A weak shoulder centered at around 5738~eV in the absorption spectra of $x =0.04$ is due to a contribution from Ce ions in the $4+$ oxidation state. For $x = 0.04$, both Ce$^{3+}$ and Ce$^{4+}$ contributions are clearly visible, indicating a valence fluctuating state for this compound, which is in agreement with the broad peak in $\chi$($T$) near 40-50~K (see Fig.~\ref{MSusc1}). As expected the intensity of the Ce$^{4+}$ component for $x = 0.15$ is strongly suppressed as compared to its value for $x = 0.04$. The observed behavior reflects a crossover from a valence fluctuating state for the Ce $4f$ state to a stable trivalent state between $x = 0.04$ and $x = 0.15$. This finding is in accordance with the results obtained by our thermodynamic and transport experiments.

	\section{Summary}
	In summary, we have investigated the effects of $5d$ electron doping on the Kondo semiconducting ground state of CeFe$_2$Al$_{10}$, using magnetic, transport, and thermal properties, $\mu$SR, and elastic and inelastic neutron scattering measurements for Ce(Fe$_{1-x}$Ir$_{x}$)$_2$Al$_{10}$ ($x\leq0.15$). With increasing $x$ to 0.15, the $b$~axis parameter decreases while changes in $a$ and $c$ leave the unit cell volume almost unchanged. The low-temperature semiconducting behavior in the resistivity with a charge gap is completely suppressed for $x\ge 0.04$, and instead a metallic-like behavior appears. Inelastic neutron scattering results reveal that the spin gap of 12.5~meV for $x = 0$ decreases to 8~meV for $x = 0.04$. For $x = 0.08$, however, the spin gap is closed and the INS response is transformed into a quasi-elastic line. It has been shown that the DOS at the Fermi level is considerably changed by $5d$ electron doping, which results in the destruction of the spin-gap excitation for $x\ge 0.08$. We observe that Ce(Fe$_{1-x}$Ir$_{x}$)$_2$Al$_{10}$ undergoes a long-range AFM transition below $T_{\mathrm{N}} = 3.1(2)$~K for $x = 0.15$. Ce L$_3$-edge XANES measurements provide direct evidence for a stable trivalent Ce$^{3+}$ state in the $x = 0.15$ sample. The appearance of AFM order is in contrast to $4d$ electron doped system Ce(Fe$_{1-x}$Rh$_{x}$)$_2$Al$_{10}$ ($x<0.2$) where the lattice expands isotropically and no long-range magnetic order appears down to 2~K. We propose that the appearance of AFM magnetic ordering is not only related to $c \mhyphen f$ hybridization but that a lattice contraction along the $b$~axis plays an important role. Our results also suggest that the $c \mhyphen f$ hybridization gap may be necessary to form the AFM order with a very high T$_{\mathrm{N}}$ seen in Ce$T_2$Al$_{10}$ ($T=$~Ru and Os), but is not necessary in Ce(Fe$_{1-x}$Ir$_{x}$)$_2$Al$_{10}$, which has a lower $T_{\mathrm{N}}$.   
	
	\begin{acknowledgments}
		We gratefully acknowledge the ISIS facility for the beam time on MERLIN (RB1690067) and the Diamond light source for the beam time on B18 Experiment No.SP17953-1. DTA would like to thank the Royal Society of London for International Exchange funding between the UK and Japan and Newton Advanced Fellowship funding between UK and China. RT thanks the Indian Nanomission for a post-doctoral fellowship. Y. M and T. T acknowledge the financial supports from JSPS, grant numbers JP26400363, JP15K05180, JP16H01076, and JP21K03473.
	\end{acknowledgments}

	\bibliography{bibliography}

\section*{Supplemental Material:}

 \subsection{Single crystals of $\text{Ce(Fe$_\mathbf{{0.85}}$Ir$_\mathbf{{0.15}}$)$_\mathbf{2}$Al$_\mathbf{{10}}$}$}
	
	Single crystals of Ce(Fe$_{0.85}$Ir$_{0.15}$)$_2$Al$_{10}$ were grown by an Al self-flux method using the same procedure given in Ref.~\onlinecite{JOUR}. A photograph of two of the crystals grown, each approximately $2 \times 7~\mathrm{mm}^{2}$ in area, is shown in Fig.~\ref{SC}(a). Figure~\ref{SC}(b) shows an x-ray Laue back diffraction photograph of one of the crystals of Ce(Fe$_{0.85}$Ir$_{0.15}$)$_2$Al$_{10}$, aligned along the [00l] orientation. Well-defined Laue diffraction spots together with the four fold symmetry confirm the single crystalline nature and good quality of the crystal. 
	
	Energy dispersive x-ray spectroscopy (EDX) analysis in a Zeiss Gemini SEM500 field emission scanning electron microscope was used to check the composition and homogeneity of one of the single crystals of Ce(Fe$_{0.85}$Ir$_{0.15}$)$_2$Al$_{10}$. Five regions, each approximately $1 \times 1~\mathrm{mm}^{2}$ in area were probed. The results were averaged over each region, and collected at selected sites within each region. The atomic percentages obtained for two of the regions, that are typical for the whole study, are presented in Table~\ref{edxTable}.

\setcounter{figure}{0}	
\begin{figure}
	\makeatletter 
	\renewcommand{\thefigure}{S\@arabic\c@figure}
	\makeatother
	\centering
	\includegraphics[width=8.5cm, keepaspectratio]{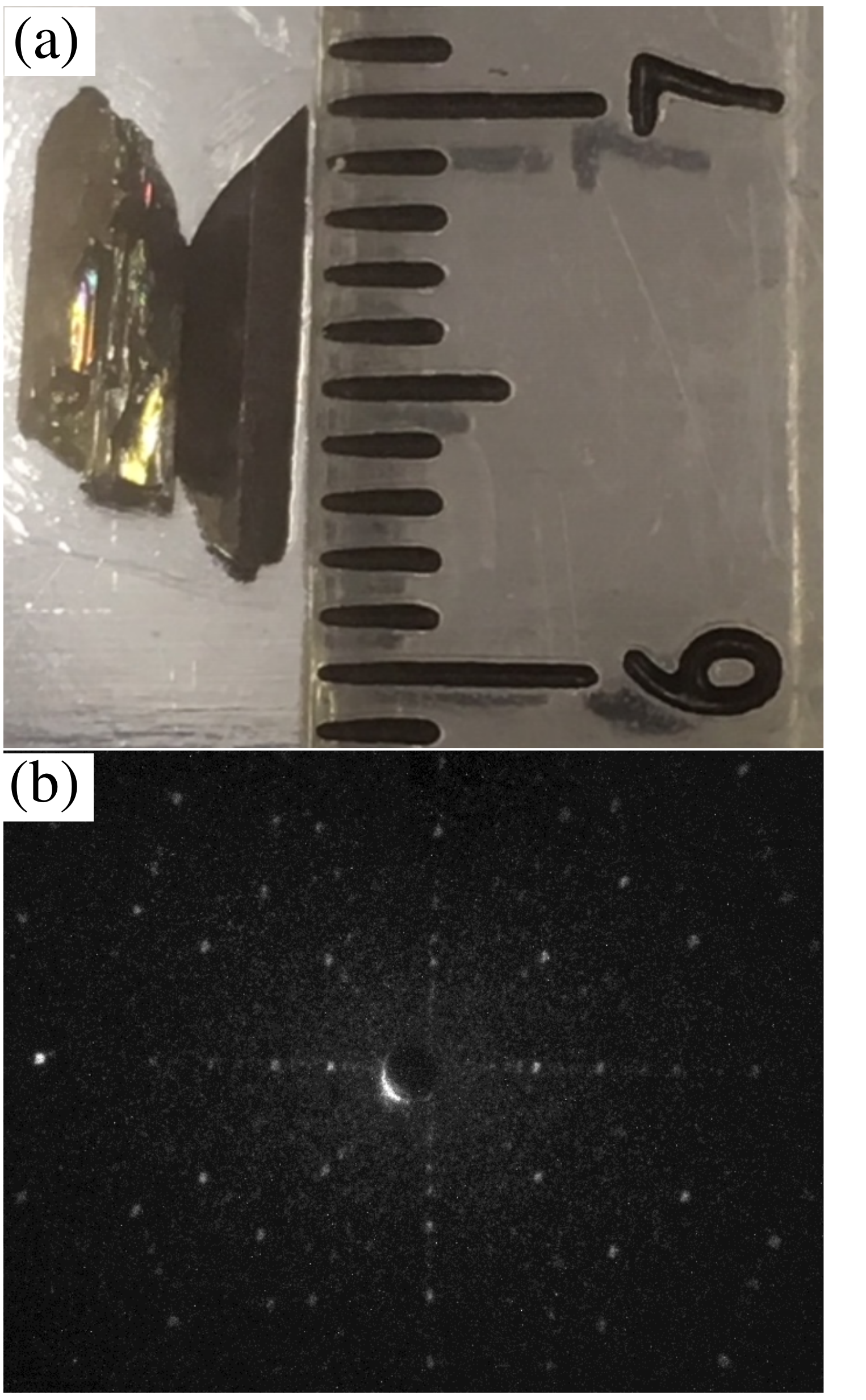}
	\caption{(a) Photograph of the two single crystals of Ce(Fe$_{0.85}$Ir$_{0.15}$)$_2$Al$_{10}$. (b) X-ray Laue pattern viewed along the $\left[0 0 1\right]$ direction for a single crystal of Ce(Fe$_{0.85}$Ir$_{0.15}$)$_2$Al$_{10}$.}
	\label{SC}
\end{figure}	
	
		\begin{table}
		\caption{Chemical composition of two of the five regions of a single crystal of Ce(Fe$_{0.85}$Ir$_{0.15}$)$_2$Al$_{10}$ studied by EDX analysis.}
		\begin{ruledtabular}
			\begin  {tabular}{ccccc }
			Spectrum  & Ce  & Fe  & Ir & Al  \\
			& (at.\%) & (at.\%) & (at.\%) & (at.\%) \\\hline 
			Region~1  &7.1(1)& 11.8(1)& 2.5(1)& 78.6(1) \\ 
			1-A  &7.1(1)& 11.6(1)& 2.4(1)& 78.9(1) \\
			1-B  &7.0(1)& 11.7(1)& 2.5(1)& 78.8(1) \\
			Region~2  &7.1(1)& 11.8(1)& 2.6(1)& 78.5(1) \\
			2-A  &7.0(1)& 11.7(1)& 2.2(1)& 79.1(1) \\ 
			2-B  &7.2(1)& 12.1(1)& 2.7(1)& 78.0(1) \\ 
			
		\end{tabular}
	\end{ruledtabular}
	\label{edxTable}
\end{table}

	\subsection{Isothermal magnetization}

\subsubsection{Polycrystalline samples of Ce(Fe$_\mathbf{{1-x}}$Ir$_\mathbf{{x}}$)$_\mathbf{2}$Al$_\mathbf{{10}}$}
\begin{figure}[tb]
		\makeatletter 
		\renewcommand{\thefigure}{S\@arabic\c@figure}
		\makeatother
		\centering
		\includegraphics[width=8.5cm, keepaspectratio]{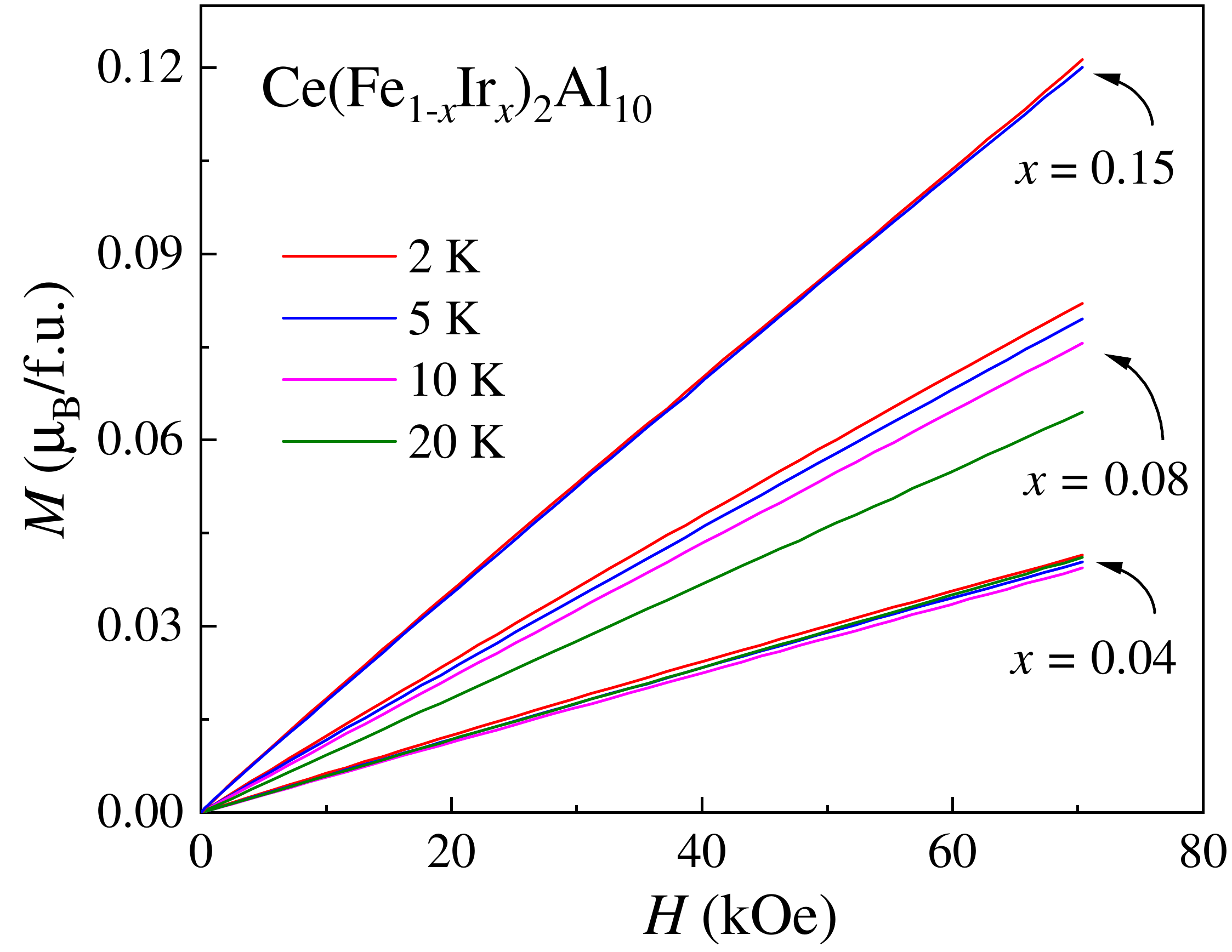}
		\caption{Magnetization versus applied field curves of polycrystalline Ce(Fe$_{1-x}$Ir$_{x}$)$_2$Al$_{10}$ for $x$ = 0.04,  0.08, and 0.15 at different temperatures.}
		\label{MH_POLY}
	\end{figure}

	Magnetization isotherms for polycrystalline samples of Ce(Fe$_{1-x}$Ir$_x$)$_2$Al$_{10}$ ($x = 0.04$, 0.08, 0.15), measured at various temperatures are displayed in Fig.~\ref{MH_POLY}. The isotherms show a linear increase with field and also increases linearly with increasing $x$. The magnetization for $x = 0.15$ (0.12~$\mu_{\mathrm{B}}$/Ce) is significantly smaller than the theoretical saturation magnetization of $gJ = 2.14~\mu_{\mathrm{B}}$ for Ce$^{3+}$ even at the highest field of $H = 70$~kOe. The magnetization data, along with a peak in the susceptibility at 3 K, hints at an antiferromagnetically ordered arrangement of the moments in the magnetically ordered state, with a reduced magnetic moment that is most likely a consequence of Kondo screening of the $4f$ moments. 
	
	\subsubsection{Single Crystal sample of Ce(Fe$_\mathbf{{0.85}}$Ir$_\mathbf{{0.15}}$)$_\mathbf{2}$Al$_\mathbf{{10}}$}

\begin{figure}[tbh]
		\makeatletter 
		\renewcommand{\thefigure}{S\@arabic\c@figure}
		\makeatother
		\centering
		\includegraphics[width=8.5cm, keepaspectratio]{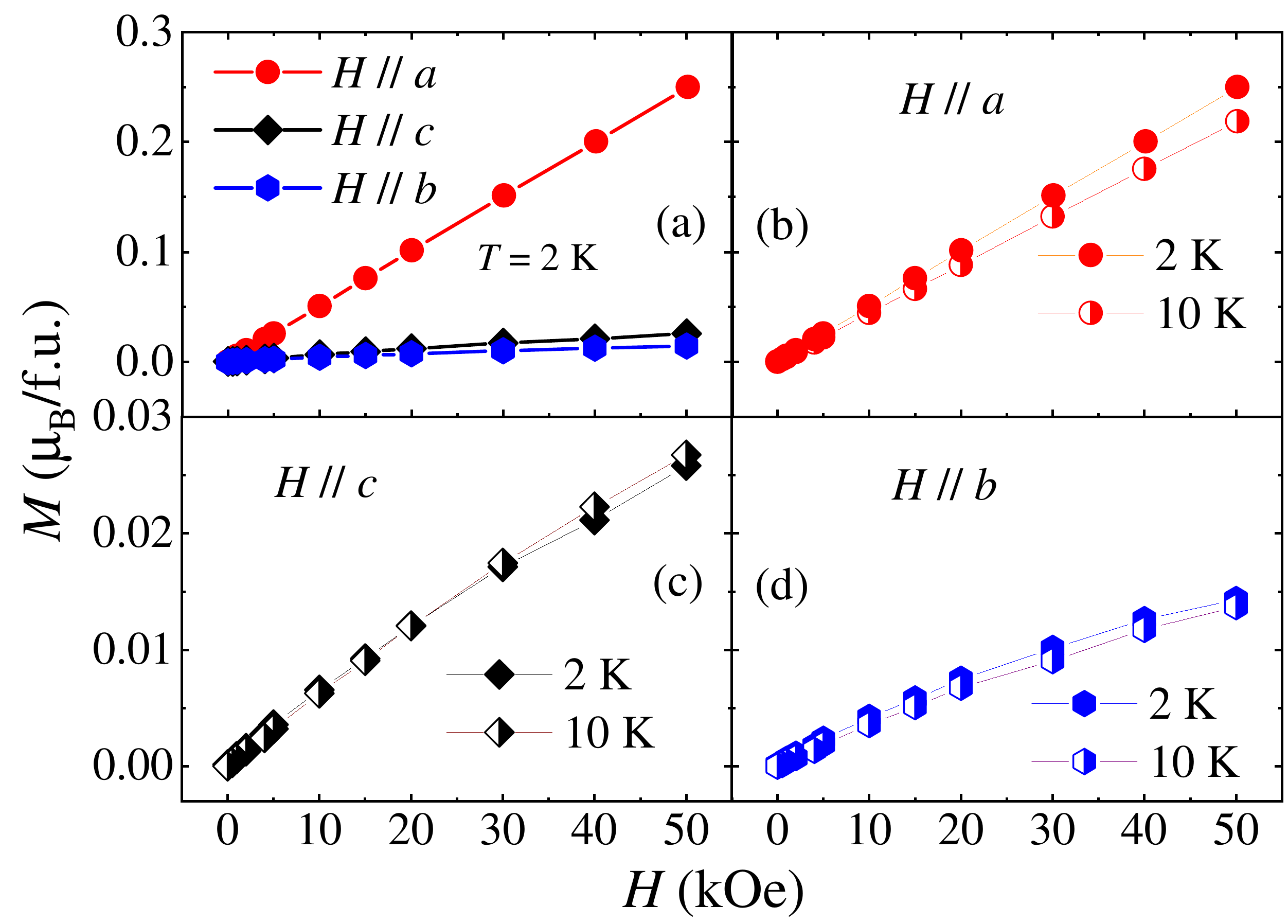}
		\caption{Isothermal magnetization of single crystal  Ce(Fe$_{0.85}$Ir$_{0.15}$)$_2$Al$_{10}$ (a) along three principal axes at 2~K, and for (b) $H \parallel a$, (c) $H \parallel c$, and (d) $H \parallel b$ at 2 and 10~K.}
		\label{MH_SC}
	\end{figure}

	Figures~\ref{MH_SC}(a)-\ref{MH_SC}(d) show the magnetization curves for Ce(Fe$_{0.85}$Ir$_{0.15}$)$_2$Al$_{10}$ with $H$ along the three principal axes at $T = 2$ and 10~K. The magnetization with $H \parallel a$ ($M_a$) is strongly enhanced over $M_b$ and  $M_c$ below $T_{\mathrm{N}}$ [Fig.~\ref{MH_SC}(a)], although the magnetic moment per Ce ion along the easy $a$ axis reaches only 0.25~$\mu_{\mathrm{B}}$/Ce at 50~kOe, which is slightly higher than for the polycrystalline sample (0.12 $\mu_{\mathrm{B}}$/Ce) but still far from the saturation value expected theoretically for Ce$^{3+}$. The $M$ versus $H$ curves increase monotonically with field up to 50~kOe with no evidence for any spin-flip transition.

	\subsection{Neutron powder diffraction study on C\lowercase{e}(F\lowercase{e}$_\mathbf{{0.85}}$I\lowercase{r}$_\mathbf{{0.15}}$)$_\mathbf{2}$A\lowercase{l}$_\mathbf{{10}}$}
	
	\begin{figure}[tbh]
		\makeatletter 
		\renewcommand{\thefigure}{S\@arabic\c@figure}
		\makeatother
		\centering
		\includegraphics[width=8.5cm, keepaspectratio]{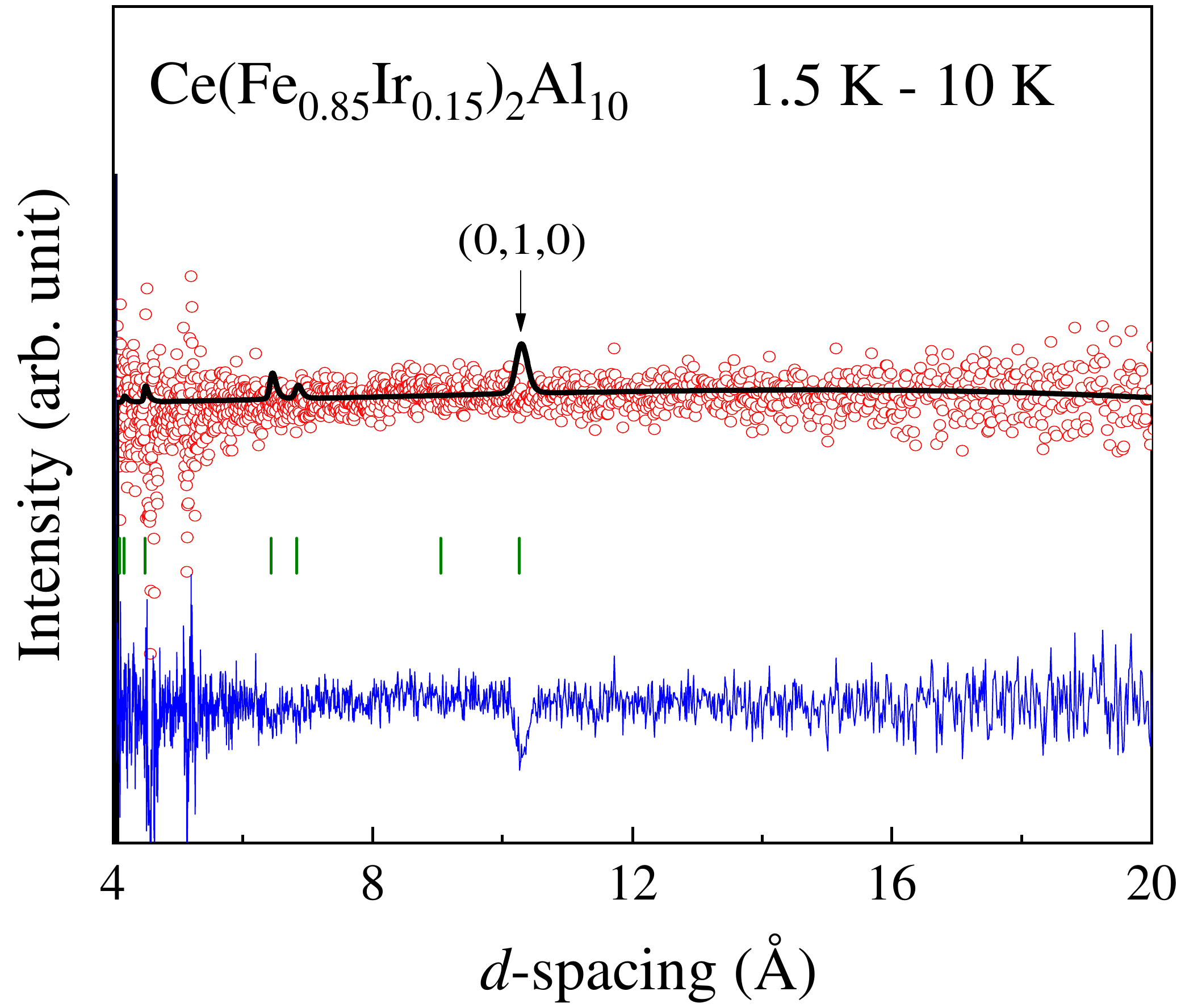}
		\caption{Powder neutron diffraction pattern obtained as a difference between the data collected at 1.5 and $10$~K. The circles (red) and solid line (black) represent the experimental and simulated (moment of 0.08~$\mu{_\mathrm{B}}$ along the $c$~axis) intensities, respectively, and the line below (blue) is the difference between them. Tick marks indicate the positions of Bragg peaks for the magnetic scattering with a $\mathbf{k}=(1,0,0)$ propagation vector.}
		\label{ND}
	\end{figure}
	
	Powder neutron diffraction measurements on Ce(Fe$_{0.85}$Ir$_{0.15}$)$_2$Al$_{10}$ were carried out at 1.5 and 10~K using the time-of-flight neutron diffractometer WISH at the ISIS Facility. Figure~\ref{ND} shows the temperature difference, 1.5~K - 10~K, diffraction pattern from the one of the detectors banks of the WISH. The temperature difference pattern does not reveal any clear evidence of magnetic ordering in Ce(Fe$_{0.85}$Ir$_{0.15}$)$_2$Al$_{10}$. A simulation of the magnetic scattering, in the models reported before for undoped and carrier-doped CeRu$_2$Al$_{10}$ and CeOs$_2$Al$_{10}$ systems, allowed us to set the upper limit for the ordered moment to be $\sim$ 0.07 $\mu_{\mathrm{B}}$/Ce in Ce(Fe$_{0.85}$Ir$_{0.15}$)$_2$Al$_{10}$. Above this value, an ordered moment in any direction would result in a statistically significant magnetic signal as shown in Fig.~\ref{ND} for the case, when the Ce-moment of 0.08~$\mu_{\mathrm{B}}$ is along the $c$~axis.

\end{document}